\documentclass[preprint,12pt]{article}

\usepackage{amsmath,amssymb,array,calc,rotating,epsfig,psfrag, amscd,graphics}

\usepackage{color}
\usepackage[
      colorlinks=true,
      urlcolor=blue,    
      filecolor=blue,     
      citecolor=red,
      pdfstartview=FitV,
       bookmarksopen=true    
      ]{hyperref}
\usepackage[left=2cm,top=1cm,right=3cm,nohead]{geometry}

\newcommand{\nc}{\newcommand}

\definecolor{cardinal}{rgb}{0.6,0,0}
\definecolor{darkgreen}{rgb}{0,0.5,0}
\definecolor{golden}{rgb}{0.92, 0.7, 0}
\definecolor{midnight}{rgb}{0, 0, 0.5}
\definecolor{darkblue}{rgb}{0.2, 0, 0.8}

\nc{\ra}{\rightarrow} 
\nc{\lra}{\leftrightarrow} 
\nc{\Ra}{\Rightarrow} 
\nc{\LRa}{\Leftightarrow} 
\nc{\blp}{{\big (}}
\nc{\brp}{{\big )}}
\nc{\Blp}{{\Big (}}
\nc{\Brp}{{\Big )}}
\nc{\bglp}{{\bigg (}}
\nc{\bgrp}{{\bigg )}}
\nc{\Bglp}{{\Bigg (}}
\nc{\Bgrp}{{\Bigg )}}
\nc{\slb}{{\rm [}}
\nc{\srb}{{\rm ]}}
\nc{\bslb}{{\rm \big [}}
\nc{\bsrb}{{\rm \big ]}}
\nc{\Bslb}{{\rm \Big [}}
\nc{\Bsrb}{{\rm \Big ]}}

\def\al{\alpha}

\def\ep{\epsilon}
\def\eps{\epsilon}
\nc{\veps}{\varepsilon}
\def\gam{\gamma}

\def\lam{\lambda}
\def\om{\omega}
\def\ph{\phi}

\nc{\vphi}{\varphi}
\def\tha{\theta}

\def\sig{\sigma}

\def\Gam{\Gamma}
\def\Ga{\Gamma}
\def\Lam{\Lambda}
\def\Om{\Omega}
\def\Sig{\Sigma}

\def\eql{=}
\def\coeff#1#2{\relax{\textstyle {#1 \over #2}}\displaystyle}

\def\ff{{\mathfrak{f}}}

\nc{\myvspace}{\rule[-1em]{0pt}{2.5em}}
\nc{\bea}{\begin{eqnarray}}
\nc{\eea}{\end{eqnarray}}
\nc{\be}{\begin{equation}}
\nc{\ee}{\end{equation}}
\nc{\barr}{\begin{array}}
\nc{\earr}{\end{array}}

\nc{\cA}{{\cal A}}
\nc{\cB}{ \cal B}

\def\cD{{\cal D}}

\nc{\cF}{{\cal F}}
\nc{\cG}{{\cal G}}

\nc{\cL}{{\cal L}}
\nc{\cM}{{\cal M}}
\def\N{{\cal N}}
\def\cN{{\cal N}}

\def\cO{{\cal O}}

\nc{\cQ}{{\cal Q}}
\nc{\cR}{{\cal R}}

\def\cV{{\cal V}}
\def\cV{{\cal V}}

\def\cZ{{\cal Z}}
\nc{\cQd}{\cQ^{\dagger}}
\nc{\cRd}{\cR^{\dagger}}
\nc{\BB}{{\mathbb B}}
\nc{\CC}{{\mathbb C}}
\nc{\DD}{{\mathbb D}}
\nc{\EE}{{\mathbb E}}
\nc{\FF}{{\mathbb F}}
\nc{\GG}{{\mathbb G}}
\nc{\HH}{{\mathbb H}}
\nc{\JJ}{{\mathbb J}}
\nc{\RR}{{\mathbb R}}
\nc{\PP}{{\mathbb P}}
\nc{\QQ}{{\mathbb Q}}
\nc{\ZZ}{{\mathbb Z}}
\nc{\calone}{{\mathbb 1}}

\nc{\half}{\frac{1}{2}}
\nc{\qrt}{\frac{1}{4}}
\nc{\del}{\partial}

\nc{\delbar}{\bar\partial}
\nc{\thalf}{\frac{t}{2}}
\nc{\Spin}{\operatorname{Spin}}
\nc{\SO}{\operatorname{SO}}

\nc{\Sp}{{\rm Sp}}
\nc{\com}[2]{{ \left[ #1, #2 \right] }}
\nc{\acom}[2]{{ \left\{ #1, #2 \right\} }}
\nc{\rr}{\rightarrow}
\nc{\p}{\partial}
\nc{\LT}{{\LL_\T}}
\nc{\Tr}{{\rm Tr}}
\nc{\tr}{{\rm tr}}
\nc{\Adag}{A^{\dagger}}
\nc{\AdagI}{A^{\dagger I}}
\nc{\AdagJ}{A^{\dagger J}}
\nc{\AdagK}{A^{\dagger K}}
\nc{\AdagL}{A^{\dagger L}}
\nc{\AdagM}{A^{\dagger M}}
\nc{\Bdag}{B^{\dagger}}
\nc{\BdagI}{B^{\dagger}_I}
\nc{\BdagJ}{B^{\dagger}_J}
\nc{\BdagK}{B^{\dagger}_K}
\nc{\BdagL}{B^{\dagger}_L}
\nc{\BdagM}{B^{\dagger}_M}
\nc{\Cdag}{C^{\dagger}}
\nc{\CdagI}{C^{\dagger I}}
\nc{\CdagJ}{C^{\dagger J}}
\nc{\CdagK}{C^{\dagger K}}
\nc{\Ddag}{D^{\dagger}}
\nc{\DdagI}{D^{\dagger I}}
\nc{\DdagJ}{D^{\dagger J}}
\nc{\DdagK}{D^{\dagger K}}
\nc{\ttha}{\tilde{\theta}}
\nc{\ttau}{\tilde{\tau}}
\nc{\tTha}{\tilde{\Theta}}
\nc{\tphi}{\widetilde{\phi}}
\nc{\tsig}{\tilde{\sig}}
\nc{\tom}{\widetilde{\om}}
\nc{\tOm}{\widetilde{\Om}}
\nc{\tlam}{\widetilde{\lam}}
\nc{\tLam}{\tilde{\Lam}}
\nc{\tSig}{\widetilde{\Sig}}
\nc{\tPhi}{\tilde{\Phi}}
\nc{\tPhibar}{\ol{\tPhi}}
\nc{\tPi}{\widetilde{\Pi}}
\nc{\tpsi}{\widetilde{\psi}}
\nc{\tPsi}{\tilde{\Psi}}
\nc{\tgam}{\widetilde{\gam}}
\nc{\tGam}{\widetilde{\Gam}}
\nc{\tzeta}{\tilde{\zeta}}
\nc{\tZeta}{\tilde{\Zeta}}
\nc{\teta}{\widetilde{\eta}}
\nc{\teps}{\tilde{\eps}}
\nc{\tveps}{\tilde{\veps}}
\nc{\tEta}{\tilde{\Eta}}
\nc{\tchi}{\tilde{\chi}}
\nc{\tChi}{\tilde{\Chi}}
\nc{\txi}{\tilde{\xi}}
\nc{\tXi}{\widetilde{\Xi}}
\nc{\tnu}{\tilde{\nu}}
\nc{\tmu}{\tilde{\mu}}

\nc{\tb}{\tilde b}
\nc{\tc}{\tilde c}
\nc{\te}{\tilde e}
\nc{\tf}{\tilde f}
\nc{\tg}{\tilde g}
\nc{\ti}{\tilde i}
\nc{\tj}{\tilde j}
\nc{\tk}{\tilde k}
\nc{\tl}{\tilde l}
\nc{\tm}{\tilde m}
\nc{\tn}{\tilde n}
\nc{\tp}{\widetilde{p}}
\nc{\tq}{\widetilde{q}}
\nc{\ts}{{\tilde s}}
\nc{\tu}{{\tilde u}}
\nc{\tv}{{\tilde v}}
\nc{\tw}{{\tilde w}}
\nc{\tx}{{\tilde x}}
\nc{\ty}{{\tilde y}}
\nc{\tz}{\tilde z}
\nc{\tA}{{\widetilde A}}
\nc{\tAbar}{{\ol \tA}}
\nc{\tB}{{\widetilde B}}
\nc{\tC}{{\widetilde C}}
\nc{\tD}{{\widetilde D}}
\nc{\tE}{{\widetilde E}}
\nc{\tF}{{\widetilde F}}
\nc{\tG}{{\widetilde G}}
\nc{\tH}{{\widetilde H}}
\nc{\tJ}{{\widetilde J}}
\nc{\tJbar}{{\ol {\tilde J}}}
\nc{\tK}{{\widetilde K}}
\nc{\tL}{{\widetilde L}}
\nc{\tM}{{\widetilde M}}
\nc{\tN}{{\widetilde N}}
\nc{\tcN}{{\widetilde \cN}}
\nc{\tP}{{\widetilde P}}
\nc{\tQ}{{\widetilde Q}}
\nc{\tR}{{\widetilde R}}
\nc{\tS}{\widetilde{S}}
\nc{\tcF}{\tilde{{\cal F}}}
\nc{\tX}{\widetilde{X}}
\nc{\tY}{\widetilde{Y}}
\nc{\tcZ}{\tilde{\cZ}}
\nc{\tcZbar}{\ol{\tcZ}}

\nc{\ha}{\hat a}
\nc{\hb}{\hat b}
\nc{\hc}{\widehat c}
\nc{\hd}{\widehat d}
\nc{\he}{\widehat e}
\nc{\hf}{\widehat f}
\nc{\hg}{\widehat g}
\nc{\hh}{\widehat h}
\nc{\hp}{\widehat p}
\nc{\hr}{\widehat r}
\nc{\hs}{\widehat s}
\nc{\hv}{\widehat v}
\nc{\hw}{\widehat w}
\nc{\hx}{\widehat x}
\nc{\hy}{\widehat y}
\nc{\hz}{\widehat z}
\nc{\zhat}{\hat z}
\nc{\hA}{\widehat{A}}
\nc{\hB}{\widehat{B}}
\nc{\hC}{\widehat{C}}
\nc{\hD}{\widehat{D}}
\nc{\hE}{\widehat{E}}
\nc{\hF}{\widehat{F}}
\nc{\hG}{\widehat{G}}
\nc{\hH}{\widehat{H}}
\nc{\hJ}{\widehat{J}}
\nc{\hK}{\widehat{K}}
\nc{\hL}{\widehat{L}}
\nc{\hM}{\widehat M}
\nc{\hN}{\widehat{N}}
\nc{\hO}{\widehat{O}}
\nc{\hP}{\widehat{P}}
\nc{\hQ}{\widehat{Q}}
\nc{\hR}{\widehat{R}}
\nc{\hS}{\widehat{S}}
\nc{\hT}{\widehat{T}}
\nc{\hU}{\widehat{U}}
\nc{\hV}{\widehat V}
\nc{\hcV}{\widehat \cV}
\nc{\hX}{\widehat X}

\nc{\heta}{\widehat{\eta}}
\nc{\hal}{\widehat \alpha}
\nc{\hphi}{\widehat{\phi}}
\nc{\hkap}{\hat{\kappa}}
\nc{\hchi}{\widehat{\chi}}
\nc{\hpsi}{\widehat{\psi}}
\nc{\hgam}{\widehat{\gam}}
\nc{\hPhi}{\hat{\Phi}}
\nc{\hPsi}{\hat{\Psi}}
\nc{\hGam}{\hat{\Gam}}
\nc{\omhat}{\widehat{\om}}
\nc{\htha}{\hat{\tha}}

\nc{\w}{\wedge}

\nc{\vb}{\vec b}
\nc{\vc}{\vec c}
\nc{\vd}{\vec d}
\nc{\ve}{\vec e}
\nc{\vf}{\vec f}
\nc{\vg}{\vec g}
\nc{\vh}{\vec h}
\nc{\vp}{\vec p}
\nc{\vq}{\vec q}
\nc{\vr}{\vec r}
\nc{\vs}{\vec s}
\nc{\vv}{\vec v}
\nc{\vw}{\vec w}
\nc{\vx}{\vec x}
\nc{\vy}{\vec y}
\nc{\vz}{\vec z}

\nc{\vB}{\vec B}
\nc{\vC}{\vec C}
\nc{\vD}{\vec D}
\nc{\vE}{\vec E}
\nc{\vF}{\vec F}
\nc{\vG}{\vec G}
\nc{\vH}{\vec H}
\nc{\vP}{\vec P}
\nc{\vQ}{\vec Q}
\nc{\vR}{\vec R}
\nc{\vS}{\vec S}
\nc{\vV}{\vec V}
\nc{\vW}{\vec W}
\nc{\vX}{\vec X}
\nc{\vY}{\vec Y}
\nc{\vZ}{\vec Z}

\nc{\ol}{\overline}
\nc{\abar}{\ol{a}}
\nc{\bbar}{\ol{b}}
\nc{\cbar}{\ol{c}}
\nc{\dbar}{\ol{d}}
\nc{\ebar}{\ol{e}}
\nc{\fbar}{\ol{f}}
\nc{\ibar}{\ol{\imath}}
\nc{\jbar}{\ol{\jmath}}
\nc{\kbar}{\ol{k}}
\nc{\lbar}{\ol{l}}
\nc{\mbar}{\ol{m}}
\nc{\nbar}{\ol{n}}
\nc{\pbar}{\ol{p}}
\nc{\qbar}{\ol{q}}
\nc{\rbar}{\ol{r}}
\nc{\sbar}{\ol{s}}
\nc{\ubar}{\ol{u}}
\nc{\vbar}{\ol{v}}
\nc{\wbar}{\ol{w}}
\nc{\xbar}{\ol{x}}
\nc{\ybar}{\ol{y}}
\nc{\zbar}{\ol{z}}

\nc{\Abar}{\ol{A}}
\nc{\Bbar}{\ol{B}}
\nc{\Cbar}{\ol{C}}
\nc{\Dbar}{\ol{D}}
\nc{\Ebar}{\ol{E}}
\nc{\Fbar}{\ol{F}}
\nc{\Jbar}{\ol{J}}
\nc{\Kbar}{\ol{K}}
\nc{\Lbar}{\ol{L}}
\nc{\cLbar}{\ol{\cL}}
\nc{\Mbar}{\ol{M}}
\nc{\Nbar}{\ol{N}}
\nc{\Pbar}{\ol{P}}
\nc{\Qbar}{\ol{Q}}
\nc{\Rbar}{\ol{R}}
\nc{\Sbar}{\ol{S}}
\nc{\Tbar}{\ol{T}}
\nc{\Ubar}{\ol{U}}
\nc{\Vbar}{\ol{V}}
\nc{\Wbar}{\ol{W}}
\nc{\Xbar}{{\overline X}}
\nc{\Ybar}{{\overline Y}}
\nc{\Zbar}{{\overline Z}}
\nc{\cZbar}{{\overline \cZ}}

\nc{\epsbar}{\ol{\epsilon}}
\nc{\lambar}{\ol{\lambda}}
\nc{\kapbar}{\ol{\kappa}}
\nc{\zetabar}{\ol{\zeta}}
\nc{\Zetabar}{\ol{\Zeta}}
\nc{\taubar}{\ol{\tau}}
\nc{\Taubar}{\ol{\Tau}}
\nc{\psibar}{\ol{\psi}}
\nc{\Psibar}{\ol{\Psi}}
\nc{\tpsibar}{\ol{\tpsi}}
\nc{\tPsibar}{\ol{\tPsi}}
\nc{\phibar}{\ol{\phi}}
\nc{\Phibar}{\ol{\Phi}}
\nc{\chibar}{\ol{\chi}}
\nc{\mubar}{\ol{\mu}}
\nc{\nubar}{\ol{\nu}}
\nc{\rhobar}{\ol{\rho}}
\nc{\ombar}{\ol{\om}}
\nc{\Ombar}{\ol{\Om}}
\nc{\Deltabar}{\ol{\Delta}}
\nc{\Thetabar}{\ol{\Theta}}
\nc{\xibar}{\ol{\xi}}
\nc{\Xibar}{\ol{\Xi}}

\nc{\Dthbar}{\ol{\rm D3}}

\nc{\gdot}{\dot{g}}
\nc{\pdot}{\dot{p}}
\nc{\qdot}{\dot{q}}
\nc{\rdot}{\dot{r}}
\nc{\sdot}{\dot{s}}
\nc{\tdot}{\dot{t}}
\nc{\udot}{\dot{u}}
\nc{\vdot}{\dot{v}}
\nc{\wdot}{\dot{w}}
\nc{\xdot}{\dot{x}}
\nc{\xddot}{\ddot{x}}
\nc{\ydot}{\dot{y}}
\nc{\zdot}{\dot{z}}
\nc{\yddot}{\ddot{y}}
\nc{\phidot}{\dot{\phi}}
\nc{\psidot}{\dot{\psi}}
\nc{\sinp}{s_{\phi}}
\nc{\cosp}{c_{\phi}}
\nc{\tanp}{t_{\phi}}
\nc{\spone}{s_{\phi_1}}
\nc{\cpone}{c_{\phi_1}}
\nc{\tpone}{t_{\phi_1}}
\nc{\sptwo}{s_{\phi_2}}
\nc{\cptwo}{c_{\phi_2}}
\nc{\tptwo}{t_{\phi_2}}
\nc{\spth}{s_{\phi_3}}
\nc{\cpth}{c_{\phi_3}}
\nc{\tpth}{t_{\phi_3}}
\nc{\calp}{c_{\al}}
\nc{\salp}{s_{\al}}

\nc{\csch}{{\rm csch}}
\nc{\sech}{{\rm sech}}

\nc{\cothzlami}{\coth(z-\lam_i)}
\nc{\coshzlami}{\cosh(z-\lam_i)}
\nc{\sinhzlami}{\sinh(z-\lam_i)}

\nc{\cothzlamj}{\coth(z-\lam_j)}
\nc{\coshzlamj}{\cosh(z-\lam_j)}
\nc{\sinhzlamj}{\sinh(z-\lam_j)}

\nc{\cothlamij}{\coth(\lam_i-\lam_j)}
\nc{\coshlamij}{\cosh(\lam_i-\lam_j)}
\nc{\sinhlamij}{\sinh(\lam_i-\lam_j)}

\nc{\bah}{{\mathbf {\hat{A}}}}
\nc{\bX}{{\mathbf X}}
\nc{\ba}{{\bf a}}
\nc{\bb}{{\bf b}}
\nc{\bc}{{\bf c}}
\nc{\bd}{{\bf d}}
\nc{\bg}{{\bf g}}
\nc{\bk}{{\bf k}}
\nc{\bl}{{\bf l}}
\nc{\bm}{{\bf m}}
\nc{\bn}{{\bf n}}
\nc{\bo}{{\bf o}}
\nc{\bp}{{\bf p}}
\nc{\bq}{{\bf q}}
\nc{\br}{{\bf r}}
\nc{\bs}{{\bf s}}
\nc{\bt}{{\bf t}}
\nc{\bu}{{\bf u}}
\nc{\bv}{{\bf v}}
\nc{\bw}{{\bf w}}
\nc{\bx}{{\bf x}}
\nc{\by}{{\bf y}}
\nc{\bz}{{\bf z}}
\nc{\bom}{{\bf \om}}
\nc{\bombar}{{\mathbf \ombar}}
\nc{\bPhi}{{\bf \Phi}}

\nc{\rma}{{\rm a}}
\nc{\rmb}{{\rm b}}
\nc{\rmc}{{\rm c}}
\nc{\rmd}{{\rm d}}
\nc{\rmg}{{\rm g}}
\nc{\rk}{{\rm k}}
\nc{\rml}{{\rm l}}
\nc{\rmm}{{\rm m}}
\nc{\rmn}{{\rm n}}
\nc{\rmo}{{\rm o}}
\nc{\rmp}{{\rm p}}
\nc{\rmq}{{\rm q}}
\nc{\rmr}{{\rm r}}
\nc{\rms}{{\rm s}}
\nc{\rmt}{{\rm t}}
\nc{\rmu}{{\rm u}}
\nc{\rmv}{{\rm v}}
\nc{\rmw}{{\rm w}}
\nc{\rmx}{{\rm x}}
\nc{\rmy}{{\rm y}}
\nc{\rmz}{{\rm z}}

\nc{\dal}{\dot{\al}}
\nc{\thadot}{\dot{\tha}}
\nc{\thab}{\bar{\theta}}
\nc{\thal}{\theta^{\al}}
\nc{\thdal}{\bar{\theta}^{\dal}}

\nc{\thsigthm}{\tha \sigma^m \thab}
\nc{\thsigthn}{\tha \sigma^n \thab}

\nc{\Dal}{D_{\al}}
\nc{\Ddal}{\bar{D}_{\dal}}
\nc{\CDal}{{\cal D}_{\al}}
\nc{\CDdal}{\bar{\cal D}_{\dal}}

\nc{\eq}[1]{(\ref{#1})}
\nc{\non}{\nonumber}
\nc{\Fzero}{F_{(0)}}
\nc{\Ftwo}{F_{(2)}}
\nc{\Ffour}{F_{(4)}}
\nc{\Fone}{F_{(1)}}
\nc{\Fthree}{F_{(3)}}
\nc{\Ffive}{F_{(5)}}
\nc{\Fn}{F_{(n)}}
\nc{\Fp}{F_{(p)}}

\nc{\tFzero}{\tF_{(0)}}
\nc{\tFtwo}{\tF_{(2)}}
\nc{\tFfour}{\tF_{(4)}}
\nc{\tFone}{\tF_{(1)}}
\nc{\tFthree}{\tF_{(3)}}
\nc{\tFfive}{\tF_{(5)}}
\nc{\tFn}{\tF_{(n)}}
\nc{\tFp}{\tF_{(p)}}

\nc{\Czero}{C_{(0)}}
\nc{\Ctwo}{C_{(2)}}
\nc{\Cfour}{C_{(4)}}
\nc{\Cone}{C_{(1)}}
\nc{\Cthree}{C_{(3)}}
\nc{\Cfive}{C_{(5)}}
\nc{\Cn}{C_{(n)}}

\nc{\equ}{{\rm eq}}

\nc{\vol}{{\rm vol}}
\nc{\Ainf}{A_{\infty}}
\nc{\End}{{\rm End}}
\nc{\Ext}{{\rm Ext}}
\nc{\IIB}{{\rm IIB}}
\nc{\Ad}{{\rm Ad}}
\nc{\IIA}{{\rm IIA}}
\nc{\AdS}{{\rm AdS}}
\nc{\CFT}{{\rm CFT}}
\nc{\Dslash}{\ensuremath \raisebox{0.025cm}{\slash}\hspace{-0.32cm} D}
\nc{\cDslash}{\ensuremath \raisebox{0.025cm}{\slash}\hspace{-0.32cm} \cD}
\nc{\omslash}{\om\!\!\!/}
\nc{\no}{\!:\!\!}
\nc{\ointdz}{\oint\frac{dz}{2\pi i}}
\nc{\ointdzone}{\oint\frac{dz_1}{2\pi i}}
\nc{\ointdztwo}{\oint\frac{dz_2}{2\pi i}}
\nc{\ointdzb}{\oint\frac{d\zbar}{2\pi i}}
\nc{\ointdzbone}{\oint\frac{d\zbar_1}{2\pi i}}
\nc{\ointdzbtwo}{\oint\frac{d\zbar_2}{2\pi i}}
\nc{\dz}{\frac{dz}{2\pi i}}
\nc{\dzb}{\frac{d\zbar}{2\pi i}}
\nc{\bpm}{\begin{pmatrix}}
\nc{\epm}{\end{pmatrix}}
 \nc{\bitem}{\begin{itemize}}
 \nc{\eitem}{\end{itemize}}
 \nc{\exercise}{\vskip 2mm \noindent {\bf Exercise:}}
 \nc{\definition}{\vskip 2mm \noindent {\bf Definition:}}
 
\def\Neql#1{{\cal N}\!=\!{#1}}
\def\coeff#1#2{\relax{\textstyle {#1 \over #2}}\displaystyle}
\def\IR{\mathbb{R}}
\def\Hfn{H}
\def\Sfn{S}

\begin{document}

\vspace{0.5cm}
\begin{center}
\baselineskip=13pt {\LARGE \bf{On Supersymmetric Flux Solutions \\ of M-theory\\}}
 \vskip1.5cm 
Nick Halmagyi,$^{*}$ Krzysztof Pilch$^{\dagger}$ and Nicholas P. Warner$^{\dagger}$   \\ 
\vskip 10mm
$^{*}$\textit{Laboratoire de Physique Th\'eorique et Hautes Energies,\\
Universit\'e Pierre et Marie Curie, CNRS UMR 7589, \\
F-75252 Paris Cedex 05, France}\\
\vskip 5mm
$^{\dagger}${\it Department of Physics and Astronomy, \\
University of Southern California,}
\centerline{\it Los Angeles, CA 90089-0484, USA}

\vskip0.5cm

\end{center}

\begin{abstract}
Motivated by the geometric structures of supersymmetric holographic RG-flows, we scan for $N=2$ $AdS_4$ solutions in M-theory. One particularly well understood holographic RG flow in M-theory is dual to a mass deformation of the $\N=8$ Chern-Simons theory. We utilize an Ansatz which is a natural generalization of this background in our scan. We find a single new solution with non-trivial internal flux and the topology of $S^7$. Interestingly, despite our Ansatz being quite general, within our system we rule out solutions with internal flux on more general Sasaki-Einstein seven manifolds.
\end{abstract}

\section{Introduction}

The study of eleven dimensional supergravity provided many of the early insights into Kaluza-Klein theory and this has taken on a new significance through applications in  holography \cite{Maldacena:1997re, Witten:1998qj, Gubser:1998bc}. Of particular interest are supersymmetric solutions that contain an explicit $AdS_{d+1}$ factor with $1\leq d\leq 6$, since each of these are expected to  provide  holographic duals of supersymmetric ground states in superconformal field theories.

Our understanding of holographic field theories in three dimensions was taken to a much higher level by fundamental developments on the gauge theory side \cite{Bagger:2007jr, Gustavsson2009a, Aharony:2008ug} and this stimulated development and analysis of  gravity duals in M theory.  In particular,  this provided an even greater impetus to  understand better the landscape of $AdS_4$ solutions in eleven-dimensional supergravity.  Our purpose here is to analyze  a new potential class of $AdS_4$ solutions with fluxes and exhibit a new such supergravity solution with $\Neql{2}$ supersymmetry.

Specifically,  we study solutions of eleven-dimensional supergravity in which the metric takes the form:
\begin{equation}
ds_{11}^2 = H^{2/3}ds_{AdS_4}^2+H^{-1/3}ds_{M_7}^2
\end{equation}
and the warp factor $H=H(y^i)$ is a function of the co-ordinates, $y^i$, on the internal seven-manifold. We will allow for  a non-vanishing membrane charge but we take  the five-brane charge to be zero:
\begin{equation}
\int_{M_7} * F^{(4)}\sim N\,,\qquad\qquad \int_{K_4} F^{(4)}=0 \,, \label{F4N}
\end{equation}
for all four cycles $K_4\subset M_7$. 

The prototypical class of such solutions are due to Freund and Rubin \cite{Freund:1980xh}
\begin{eqnarray}
ds_{11}^2&=& ds_{AdS_4}^2+ 6L^2 ds_{M_7}^2\,, \\
F^{(4)}&=& \frac{3}{L} \, \vol_{AdS_4}\,,
\end{eqnarray}
where 
\begin{equation}
Ric_{4}= \frac{3}{L^2}g_{4}\,,\ \ \ Ric_{7}=\frac{3}{2L^2}g_{7}\,.
\end{equation}
To preserve supersymmetry in a generic Freund-Rubin solution (as opposed to the special $S^7$ solution) there are additional geometric constraints.  In particular,  for $\frac{1}{4}$-supersymmetry to be preserved, the internal manifold $M_7$ must be Sasaki-Einstein ($SE_7$) and the dual superconformal field theory in three dimensions has $\Neql{2}$ supersymmetry. Seven dimensional,  homogeneous, regular Sasaki-Einstein manifolds have been classified \cite{Castellani:1983yg, Duff:1986hr} and the five such manifolds are usually labelled:
\begin{equation}
S^7\,, M^{3,2}\,, Q^{1,1,1}\,,N^{1,1}\ \ {\rm and} \ \ V^{5,2}  \,.
\end{equation}
An infinite class of  non-homogeneous but explicitly known Sasaki-Einstein metrics (referred to here as the GMSW solutions) was constructed in \cite{Gauntlett:2004hh} and this class of explicit $SE_7$ metrics was  further enlarged in  \cite{Chen:2004nq}.

A natural generalization of the Freund-Rubin solutions is to allow for $F^{(4)}$ to be  non-vanishing  on the internal manifold, $M_7$. We  will consider such solutions here but our solutions will only carry dielectric five-brane charge, that is, they will not have a net five-brane charge as implied by \eq{F4N}. In this sense we are looking for solutions that correspond to the near horizon limit of membranes in eleven dimensions. Somewhat surprisingly, there is a distinct paucity of explicit examples of this form in the current literature. The existing solutions are related to uplifts of critical points of four dimensional, $\Neql8$ gauged supergravity \cite{Warner:1983vz, Fischbacher:2009cj,Fischbacher:2010ec, Fischbacher:2011jx}.  Thanks to the proof that the $\N=8$ theory is a consistent truncation of eleven dimensional supergravity \cite{deWit:1984nz, deWit:1986mz, deWit:1986iy, Nicolai:2011cy} one can claim that any critical point of the seventy-dimensional scalar potential lifts to a solution of the $d=11$ theory although, in practice, performing this lift explicitly can be prohibitively difficult.

Only one non-trivial critical point of the $\Neql8$ theory preserves $\frac{1}{4}$-supersymmetry  \cite{Warner:1983vz, Nicolai:1985hs} and this solution is characterized by the fact that it preserves $SU(3)\times U(1)$ subspace of the $\subset SO(8)$ global symmetries. The lift of the $SU(3)\times U(1)$ invariant fixed point was performed in \cite{Corrado:2001nv} along with the entire supersymmetric domain wall \cite{Ahn:2000aq, Ahn:2000mf} that connects this critical point to the critical point at the origin. The lift of the $SU(3)\times U(1)$ invariant fixed point (referred to here as the CPW solution), is interesting in its own right as a solution of the eleven-dimensional theory as it is the first explicit example of a supersymmetric $AdS_4$ solution with an internal flux but only electric membrane charge.

The internal manifold, $M_7$, of the CPW solution has the topology of $S^7$ which, for our purposes, is best viewed as a fibration of $S^5\times S^1$ over an interval $I$. The group of isometries of the metric is $SU(3)\times U(1)^2\subset SO(8)$ but the four form flux is charged under one of the $U(1)$'s and thus breaks this symmetry group to $SU(3)\times U(1)$. Interestingly, one can show that, for the CPW solution, the components of $F^{(4)}$ on the internal manifold  are proportional to the holomorphic two form (and its complex conjugate) on $\CC\PP^2\subset S^5$  and this is the origin of the breaking of the $U(1)$ symmetry of the Hopf fiber of $S^5$. The $U(1)$ charge of the four-form flux arises in much the same manner as it does in the non-supersymmetric solutions of \cite{Pope:1984bd, Pope:1984jj} where charge of the holomorphic three-form on $\CC\PP^3\subset S^7$ is required for global regularity.

Our purpose here is to use the structure of the CPW solution as a template for more general solutions. Indeed, it was observed in \cite{Corrado:2001nv} that while one can locally replace the $\CC\PP^2\subset S^5\subset S^7$ by any K\"ahler-Einstein 4-manifold and still solve the field equations this will introduce singularities in $M_7$ at a particular endpoint of the interval $I$. Nonetheless, it is conceivable that a full analysis of the BPS equations could reveal further integration constants which allow the singularities to be avoided by judicious choices of boundary conditions of $I$. This is the spirit of construction of Sasaki-Einstein metrics in \cite{Gauntlett:2004hh}, which resolves the singularities uncovered in \cite{Page1987}. Our Ansatz will be designed to follow up this observation, we consider an internal seven-manifold,  \eq{sevenmet}, which is $SE_5\times U(1)$ fibered over an interval $I$  and the three-form potential, \eq{AAnsatz},  is proportional to a one-form wedged with the the holomorphic two form inherited from the K\"ahler-Einstein four manifold inside the $SE_5$. We reduce the BPS equations for this Ansatz to a pair of O.D.E.'s
\begin{equation}\begin{split} 
{d{\Sfn}\over d\rho} & \eql 4\, {\Sfn\over \rho}-2 \,\Hfn\,{\Sfn^3\over\rho^3}\,,\\
{d\Hfn\over d\rho} & \eql 2\,(1-\Hfn^2)\,{\Sfn^4\over \rho^3(1-\Sfn^2)}\,.
\end{split}\non
\end{equation}
In these equations, $H(\rho)$ is the warp factor of the four/seven split of the eleven-dimensional metric and is also the rotation parameter in the di-electric spinor. The function, $S(\rho)$, has a more complicated but explicit relation to the rest of the Ansatz and essentially governs how the supersymmetry mixes between the internal manifold and the radial $AdS$ direction. Having derived the BPS equations, we also do some numerical analysis on the solution space, paying careful attention to  regularity.

Interestingly, our work is orthogonal to the work \cite{Petrini:2009ur, Lust:2009mb, Aharony:2010af, Tomasiello:2010zz} where large families of $\frac{1}{4}$-BPS, $AdS_4$ solutions have been constructed in IIA supergravity with non-vanishing Romans mass \cite{Romans:1985tz}.  Solutions with non-vanishing Romans mass cannot be lifted to eleven-dimensional supergravity and, conversely, our Ansatz cannot be reduced to supersymmetric solutions of IIA supergravity because the residual $U(1)$ symmetry in our Ansatz is the $\cR$-symmetry and so compactification along this $U(1)$ will break the supersymmetry.
 
This paper is organized as follows: In section 2 we outline our Ansatz and the methods we have used to reduce the BPS conditions to a pair of ODE's. In section 3 we review the known analytic solutions to this Ansatz. In section 4 we numerically solve the pair of ODE's. We reproduce the known analytic solution and produce a new solution with the topology of $S^7$.
\vskip 3mm
\noindent {\bf Note added:} While preparing this paper for submission we became aware of the work \cite{Gabella:2012rc} which may have some overlap with our paper.

\section{The Supersymmetry Conditions}

\subsection{The Background} 

While our solution space will ultimately be determined by just two functions, $\Hfn(\rho)$ and $\Sfn(\rho)$ we first outline our complete Ansatz. 
The metric is:
\begin{equation}\label{}
ds_{11}^2\eql H^{2/3}ds_{AdS_4}+6L^2H^{-1/3}ds_7^2\,,
\end{equation}
where
\begin{equation}\label{sevenmet}
ds_7^2\eql X_2^2 \, d\rho^2+\rho^2 ds_{B_2}^2+\left(X_4\, (d\psi+A)+X_5 \, d\phi\right)^2+X_6^2\, d\phi^2\,,
\end{equation}
and all functions depend only on $\rho$, which parametrizes the interval, $I$. The end-points of $I$ will be determined by regularity. The base, $B_2$, is taken to be   a Kahler-Einstein manifold, normalized by  $R_{\alpha\beta}= 6 g_{\alpha\beta}$.   This metric (\ref{sevenmet})  expresses $M_7$ as a fibration of $SE_5\times U(1)_{\phi}$ over $I$.  

To analyze the spinors,  we  choose the frames:
\begin{equation}\label{metanz}
\begin{split}
e^{1,2,3} & \eql H^{1/3}e^{r/L}dx^{0,1,2}\,,\\[6 pt]
e^4 & \eql \sqrt{1-S^2} H^{1/3}dr+\sqrt 6 L H^{-1/6}SX_2d\rho\,,\\[6 pt]
e^5  & \eql S H^{1/3}dr-\sqrt 6 L H^{-1/6}\sqrt{1-S^2} X_2d\rho\,,\\[6 pt]
e^{6,7,8,9} & \eql \sqrt 6 L H^{-1/6} \, \rho \, f^{1,2,3,4}\,,\\[6 pt]
e^{10} & \eql \sqrt 6 L H^{-1/6} \left(X_4(d\psi+A)+X_5d\phi\right)\,,\\[6 pt]
e^{11} & \eql \sqrt 6 L H^{-1/6} X_6 d\phi\,.
\end{split}
\end{equation}
The frames, $f^i$, are chosen so that the canonical holomorphic frames on $B_2$ are:  
\begin{equation}\label{}
\frak f^1\eql f^1-if^4\,,\qquad \frak f^2\eql f^2-if^3\,.
\end{equation}
Then the complex-structure and (local) holomorphic $(2,0)$-form on $B_2$ are:
\begin{equation}\label{}
j   \equiv \coeff i 2\, (\ff^1\wedge  \bar\ff^1+\ff^2\wedge\bar\ff^ 2)\,,\qquad \omega \equiv i\,\ff^1\wedge\ff^2\,,
\end{equation}
and we have:
\begin{equation}
dA\eql 2j\,,\qquad d\omega\eql 3i A\wedge \omega \,,\qquad \omega \wedge\omega^*\eql 2 j\wedge j\,.
\end{equation}
We can also define the fundamental forms of a $SE_5$ structure:
\begin{equation}
\label{SEstr}
\eta=(d\psi + A)\,,\ \ \ J=j\,,\ \ \ \Om=e^{3i\psi}\om\,,
\end{equation}
which satisfy 
\begin{eqnarray}
\Om\w \Om^*&=& 2 J\w J\,,\ \ \ \ \ d\eta=2J\,,\ \ \ \ \  dJ=0\,,\ \ \ \ \ d\Om=3i\eta\w \Om\,.
\end{eqnarray}

Following \cite{Corrado:2001nv}, we assume that the internal components of the three-form  potential are proportional to the holomorphic $(2,0)$ form, $\Omega$:
\begin{equation}\label{AAnsatz}
C^{(3)}  \eql  p_0 \,e^1\wedge e^2\wedge e^3 
~+~  {\rm Re}\left[\,  \left(a_1 \, d \rho  +  a_2    \, (d\psi+A)  + a_3  \, d \phi   \right)\wedge \Om \, \right] 
\end{equation}
where $p_0$ is a real function of $\rho$ and the $a_j$ are complex functions of $\rho$.  Note that there $AdS$-invariance implies that there can be no internal parts proportional to $dr$.  

\subsection{The Supersymmetry Projectors} 

Using techniques similar to \cite{Nemeschansky:2004yh}, we require that $\frac{1}{4}$-supersymmetry is preserved. This requires that the eleven-dimensional spinor, $\eps$, satisfies three projection conditions.  
Define
\begin{equation}\label{}
\Pi_0\eql  {1\over 2}\left({\bf 1}+\Ga^{123}\right)\,,\qquad
\Pi_1\eql {1\over 2}\left({\bf 1}+\Ga^{78}\Ga^{69}\right)\,,\qquad
\Pi_2\eql {1\over 2}\left({\bf 1}-\Ga^{78}\Ga^{5\,10}\right)\,.
\end{equation}
and the di-electric projector:
\begin{equation}\label{}
\tPi_0(\xi,\beta)\eql  {1\over 2}\left({\bf 1}+\cos\beta\,\Ga^{123}-\sin\beta\,\Ga^{123}(\cos\xi\,\Ga^{79\,10}+\sin\xi\,\Ga^{89\,10})\right)\,,
\end{equation}
The Killing spinor, $\epsilon$, that we seek satisfies:
\begin{equation}\label{}
\tPi_0(3\psi,\beta)\epsilon\eql \Pi_1\ep \eql\Pi_2\ep \eql 0\,,
\end{equation}

If one introduces the rotations 
\begin{equation}\label{rot1}
\cR(x)\eql \cos{x\over 2}-\sin{x\over 2}\,\Ga^{78}\,,
\end{equation}
and
\begin{equation}\label{rot2}
\cO(x)\eql \cos{x\over 2}+\sin{x\over 2}\,\Ga^{79\,10}\,.
\end{equation}
then the Killing spinor may be written as
\begin{equation}\label{killsp}
\ep\eql H^{1/6}e^{r/(2L)}\,\cR(3\psi)\,\cO(\beta)\,\cR(-2\ph)\,\ep_0\,. 
\end{equation}
where  $\ep_0$ is a {\it constant\/} spinor satisfying the simpler proections:
\begin{equation}\label{}
\Pi_0\ep_0\eql\Pi_1\ep_0\eql\Pi_2\ep_0\eql 0\,,
\end{equation}
The rotations (\ref{rot1}) and  (\ref{rot2})  commute with $\Pi_1$ and $\Pi_2$, but rotate with $\Pi_0$ in $\tPi_0$, and  $ \cO(\beta)$, in particular, represents a partial polarization of the $M2$-branes into dielectric $M5$-branes.

\section{The Analytic Solution Space}

\subsection{The Supersymmetry Conditions}

It is now relatively easy to solve the supersymmetry conditions based upon our Ansatz for the background (\ref{metanz}), (\ref{AAnsatz}) and the supersymmetries  (\ref{killsp}).  

We find that everything can be parametrized the two functions, $\Hfn(\rho)$ and $\Sfn(\rho)$,  that were used in the definition of our frames (\ref{metanz}).   
The remaining metric functions may then be written as 
%
\begin{equation}\label{solXs}
\begin{split}
U(\rho)^{-1/2}& ~\equiv~ X_2    \eql -\sqrt{2\over 3}\,{\sqrt \Hfn}\,{ \Sfn\over\rho\,\sqrt{1-\Sfn^2}}\,, \qquad 
X_4   \eql \sqrt{3\over 2}\,{1\over\sqrt \Hfn}\,{\rho^2\over \Sfn}\,,  \\[6 pt] 
X_5  &  \eql  -{1\over\sqrt6}\,{1\over\sqrt{\Hfn}}\,\Sfn\,, \qquad 
X_6    \eql {1\over\sqrt6}\,\sqrt \Hfn\,\sqrt{1-\Sfn^2}\,.
\end{split}
\end{equation}
We have defined $U(\rho)$ in analogy with the same function in \cite{Gauntlett:2004hh}.  The di-electric angle is given by:
\begin{equation}\label{}
\cos\beta   \eql   {1\over \Hfn} \,, \qquad \sin\beta   \eql    \Big(1   - {1\over \Hfn^2} \Big)^{\frac{1}{2}}\ \,.
\end{equation}

The three-form potential can be simplified to the form:
\begin{equation}\label{fxanzuv}
C^{(3)}  \eql  p_0 \,e^1\wedge e^2\wedge e^3 
+ 6 L^2 \rho^2 H^{-1/3}  \,{\rm Re}\left[ -i \left(\, p_1\,(e^4+ie^{11})+p_2\,(e^5+ie^{10})\right)\wedge \Om\right] \,,
\end{equation}
where we must have $\Sfn\,  p_2 =  -( \sqrt{1-\Sfn^2})\, p_1$ in order to cancel the $dr$ components.  Indeed we find:
\begin{equation}\label{}
p_0\eql  - {1\over 2}\cos\beta\,,\qquad p_1\eql  {\Sfn\over \sqrt{1-\Sfn^2}}\,{1\over 2}\,\tan{\beta\over 2}\, \,,\qquad p_2\eql -{1\over 2}\tan{\beta\over 2}\,.
\end{equation}

Finally, the remaining supersymmetry conditions reduce to the first order system on $\Hfn$ and $\Sfn$:
\begin{equation}\begin{split} \label{SHeqs}
{d{\Sfn}\over d\rho} & \eql 4\, {\Sfn\over \rho}-2 \,\Hfn\,{\Sfn^3\over\rho^3}\,,\\
{d\Hfn\over d\rho} & \eql 2\,(1-\Hfn^2)\,{\Sfn^4\over \rho^3(1-\Sfn^2)}\,.
\end{split}
\end{equation}
These equations thus completely determine our solutions.

\subsection{The GMSW Sasaki-Einstein Solutions}

The solutions of \cite{Gauntlett:2004hh} are obtained from a particularly simple set of solutions to \eq{SHeqs}
\begin{equation} \label{GSeqs}
\Hfn ~=~  1\,, \qquad   \Sfn(\rho) ~=~ \sqrt {3 \over 2}\,{\rho^4\over\sqrt{c+\rho^6}}\,.
\end{equation}
This implies that $\beta =0$ and hence there are no fluxes.  The metric function, $U(\rho)$, also becomes:
\begin{equation}\label{GUfnct}
U(\rho)\eql \,\Big(\,1-{3\over 2}\,\rho^2+{c\over\rho^6}\,\Big)\,,
\end{equation}
where $c$ is the integration constant, proportional to $\kappa$ in \cite{Gauntlett:2004hh}.  When $c=0$, $KE_4=\CC\PP^2$ and we obtain $M_7=S^7$ and $0<\rho<\sqrt{2/3}$. At $\rho=0$ the space is obviously smooth while at the rightmost endpoint, there is a bolt.

\begin{figure}[t]
\begin{center}
\includegraphics[width=3 in]{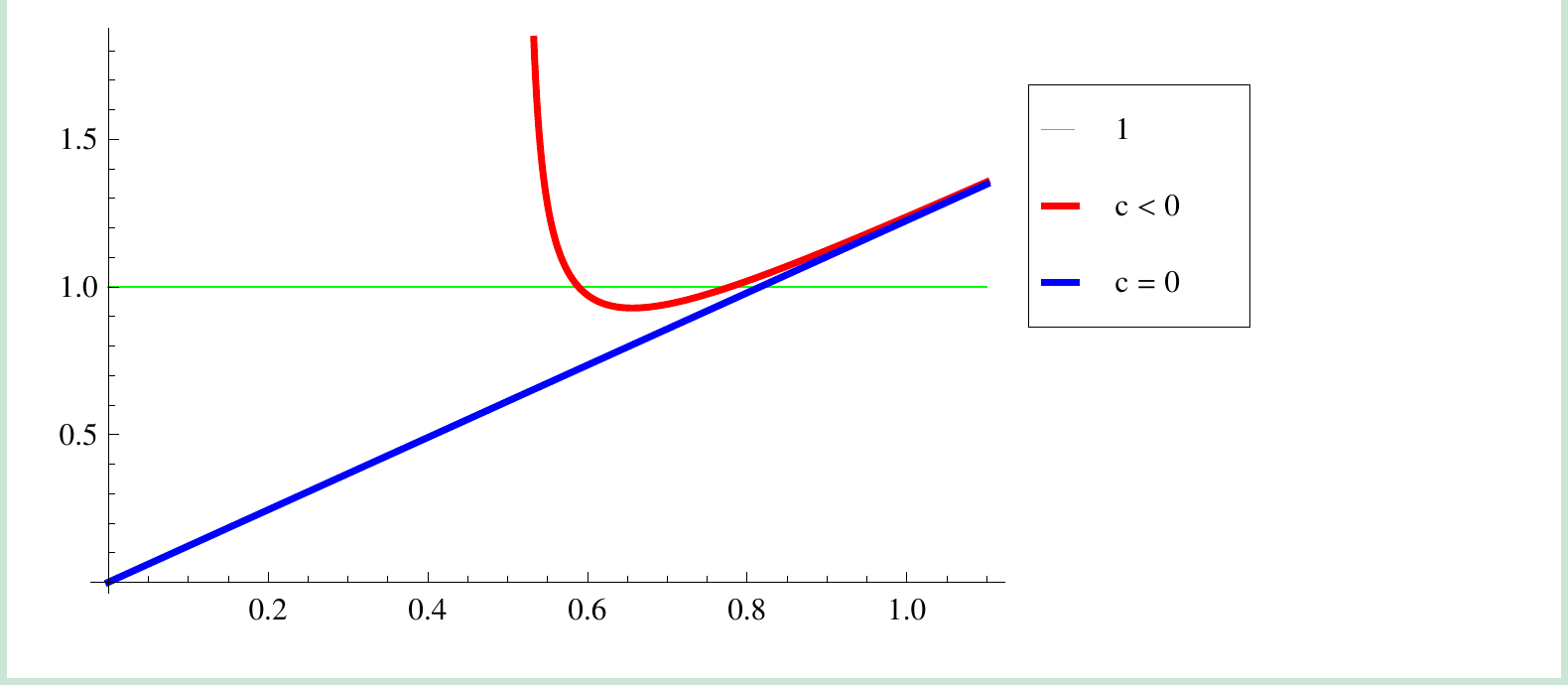}
\includegraphics[width=3 in]{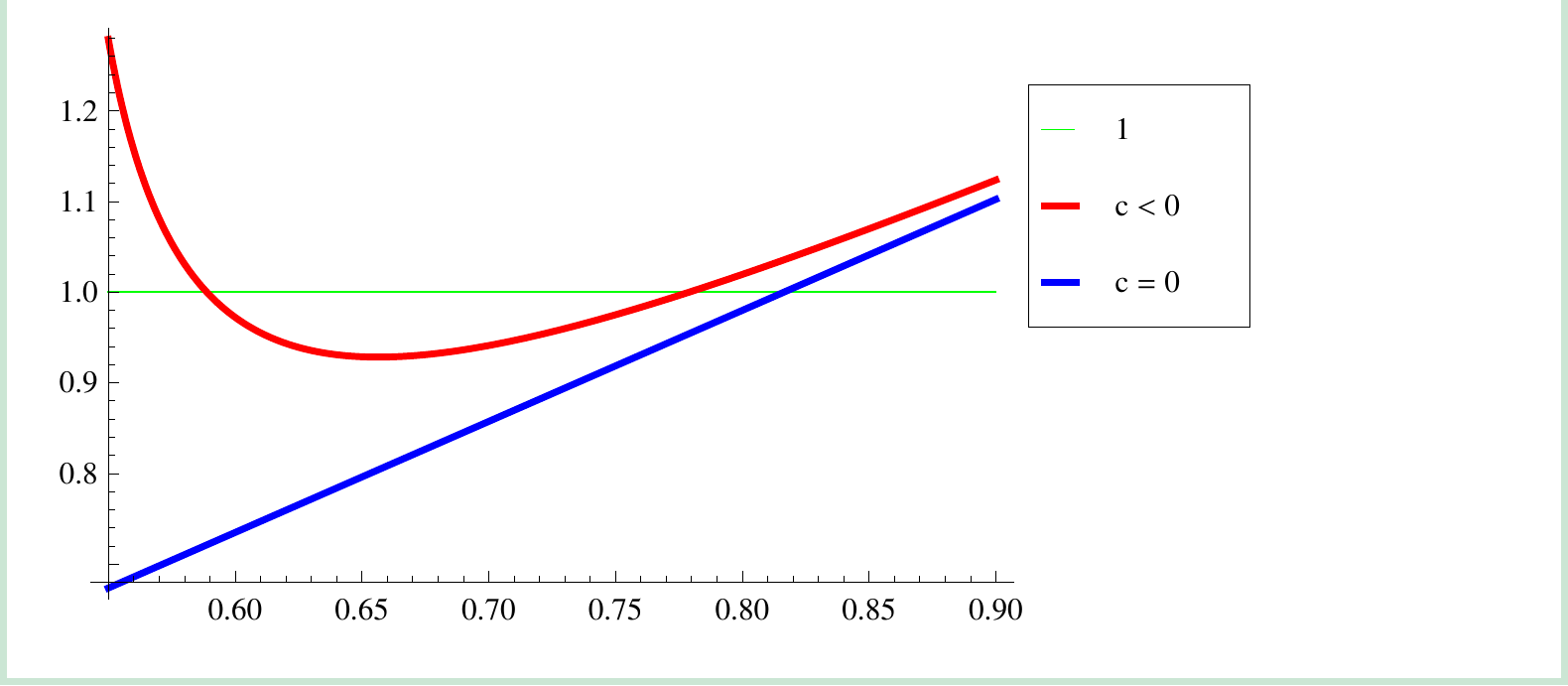}
\end{center}
\caption{\label{gntS} The {\small GMSW} solution: $S(\rho)$ for $c=0$ and $c_0<c<0$.}
\end{figure}

\begin{figure}[t]
\begin{center}
\includegraphics[width=3 in]{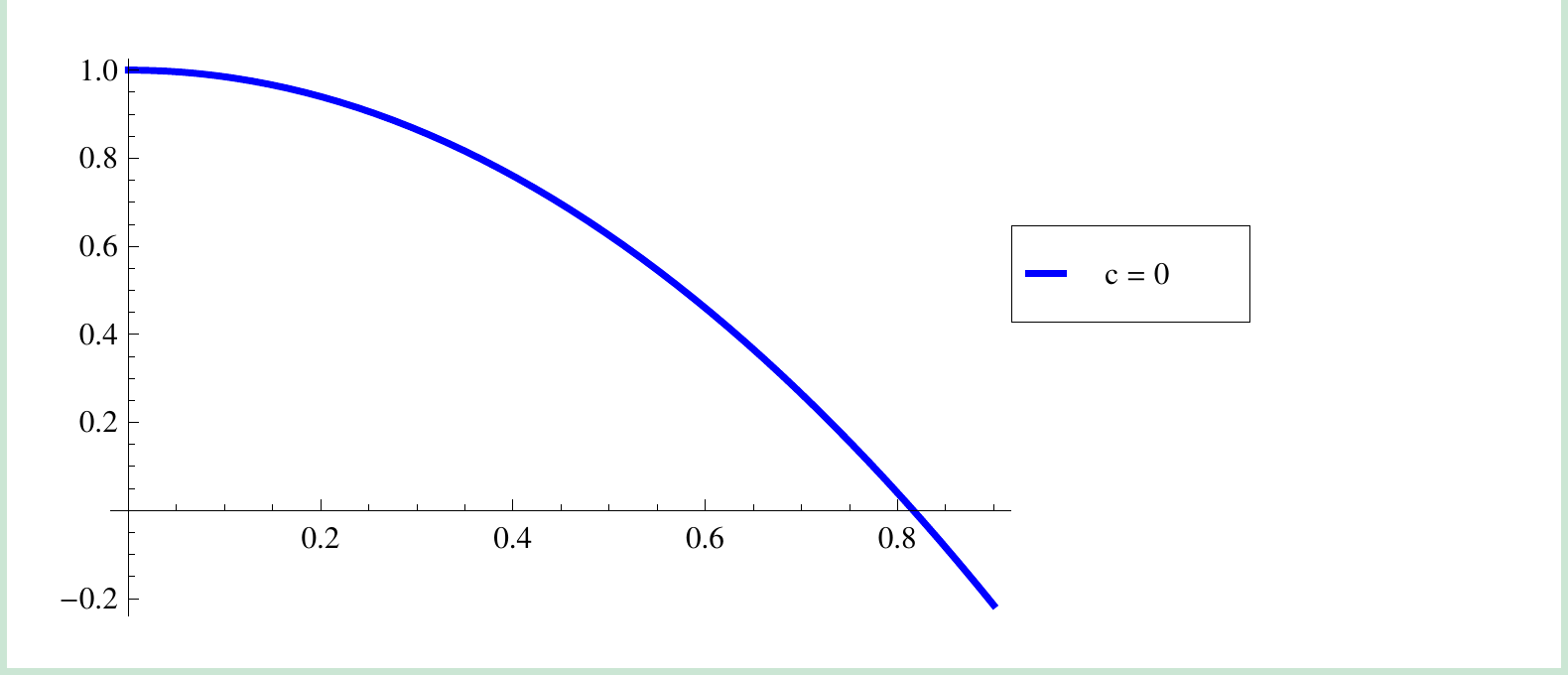}
\includegraphics[width=3 in]{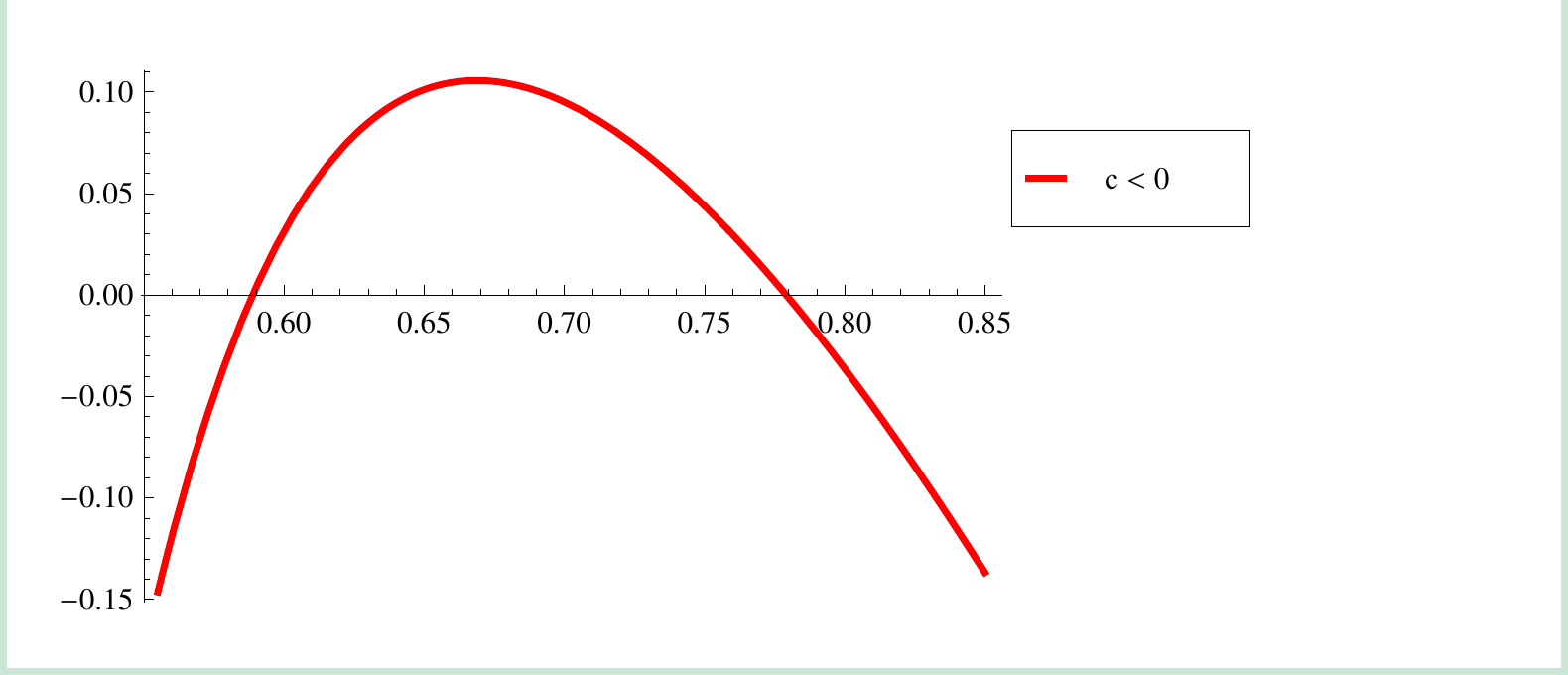}
\end{center}
\caption{\label{gntU} The {\small GMSW} solution: $U(\rho)$ for $c=0$ and for $c_0<c<0$.}
\end{figure}

The GMSW solutions have $c<0$ in the range where $U(\rho)$ is finite and positive on some interval    $0<\rho_i<\rho<\rho_{i+1}$. On that interval we must have $0<\Sfn(\rho)\leq 1$, where the equality holds only at the zeros, $\rho_i$. The plots illustrating this are in Fig.~\ref{gntS} and \ref{gntU}. Regularity of these solutions at the endpoints can be checked analytically. These solutions can be constructed upon any $KE_4$ and one finds an infinite discrete family parameterized by $c$.  


\subsection{The CPW solution}

The CPW solution  is obtained by setting
\begin{equation}\label{CPWsol}
\Hfn(\rho)\eql 3-2\rho^2\,,\qquad \Sfn\eql {\rho\over\sqrt{2-\rho^2}}\,,\qquad 0\leq\rho\leq 1\,.
\end{equation}

The solution is, of course, smooth but it is useful to examine the details in our parametrization.   At the leftmost endpoint of $I$, we have 
\begin{equation}
ds_7^2~=~ d\rho^2  ~+~ \rho^2 \Big[ ds^2_{B_2}+  \big(d\psi-\coeff{1}{6}d\phi+A\big)^2 \Big]  + \big( \coeff{1}{2}-\coeff{21}{36} \rho^2\big)\,    d\phi^2+  \cO(\rho^4)\,.
\end{equation}
First note that the $S^1$ defined by the $\phi$-circle remains large as $\rho \to 0$.  If one sets $d \phi =0$ then the remaining metric is simply a cone over the $U(1)$ Hopf fibration over $\CC \PP^2$.  If the angle, $\psi$, has period $2 \pi$ then this is a cone over $S^5$ and is thus  precisely the flat metric on $\IR^6$;   exactly as one should expect if one describes a smooth  $S^7$ as $S^5 \times S^1$ fibered over the interval, $I$.

At the rightmost endpoint of $I$, we set $ (1-\rho) = \frac{1}{4} x^2$ and find 
\begin{equation}
ds_7^2~\sim~ \coeff{1}{6} \, \big( d x^2 ~+~x^2\,   (d\phi -  \coeff{1}{2} \, \chi  )^2  \big)  +  ds^2_{B_2}+  \coeff{1}{6} (1 -  \coeff{5}{4} \,x^2\,) \, \chi^2 \,,
\end{equation}
where $\chi \equiv d \phi + 3(d\psi +A)$.  The $\CC \PP^2$ and the fiber defined by  the differential, $\chi$, now remain large as $x\to 0$.  Dropping the  $ ds^2_{B_2}$  and $\chi$ terms leaves $d x^2 + x^2 d\phi^2$, which defines a smooth $\IR^2$ if and only if $\phi$ has period $2 \pi$. Again, this is  precisely what one expects for a smooth $S^7$.  Note that the regularity of the metric determines the periods of the angular coordinates $\psi$ and $\phi$ as it did in  \cite{Gauntlett:2004hh}.

\section{The Numerical Solution Space}

The two distinct classes of known solutions to \eq{SHeqs} are classified by whether $U(\rho)$ has a zero at both endpoints of $I$ or just the rightmost endpoint. The $SE_7$ solutions are of the first kind while the round and squashed (CPW) $S^7$  solutions are of the second kind. We now investigate solutions of both kind, starting by characterizing the zeros, $\rho_i$, of $U(\rho)$.

\subsection{Expanding around a zero of $U(\rho)$}
We can tabulate the zero's of $U(\rho)$ as follows. It is clear from \eqref{SHeqs} that at such a regular zero, $\rho_i$, we must have 
\begin{equation}\label{}
\Sfn(\rho_i)=\Hfn(\rho_i)=1 \,.
\end{equation}
Using series expansions
\begin{equation}\label{}
\Sfn(\rho)\eql 1+\sum_{n\geq 1}s_n\,(\rho-\rho_i)^n\,,\qquad 
\Hfn(\rho)\eql    1+\sum_{n\geq 1}h_n\,(\rho-\rho_i)^n\,,
\end{equation}
we find that  the  constant terms in the expansion of \eqref{SHeqs} yield:
\begin{equation}\label{rec1}
s_1-{4\over \rho_i}+{2\over\rho_i^3}  \eql 0\,,\qquad h_1\,\big(1-{2\over s_1\rho_i^3}\big)\eql 0\,,
\end{equation}
while the higher terms have the structure
\begin{equation}\label{}
s_n+ \sigma_n(s_k,h_k)\eql 0\,,\qquad
h_n\,\omega_n(s_k,h_k)+\mu_n(s_k,h_k)\eql 0\,,\qquad k=1,\ldots,n-1\,,
\end{equation}
where
\begin{equation}\label{}
\mu_n(s_k,h_k)\eql 0\qquad {\rm if}\qquad h_1\eql h_2\eql\ldots\eql h_{n-1}\eql 0\,.
\end{equation}
Suppose first that $h_1\not=0$. Then the consistency of recurrence arises at the first step and \eqref{rec1} implies
\begin{equation}\label{}
\rho_0\eql 1\,,\qquad s_1\eql 2\,.
\end{equation}
The  higher order terms yield a unique solution in terms of a single parameter, $h_1$.

However, if we have $h_1=\ldots= h_i= 0$ and $h_{i+1}\not=0$, the first $i$-steps of the recurrence are trivially solved for $s_1,\ldots,s_i$ in terms of $\rho_i$. Now, the consistency condition  arises at order $i+1$ where we must solve $\omega_i\eql 0$, which is an equation for the $i$-th zero $\rho_i$. This determines the discrete set of zeros \eqref{diszer}. The higher orders are then solved without any further conditions and yield a  family of solutions  parametrized by $h_{i+1}$.

So we obtain the result that the analytic zeros of $U(\rho)$ are given by:
\begin{equation}\label{diszer}
\rho_0=1,
\qquad \rho_1\eql \frac{\sqrt{3}}{2},\ \quad \ldots\ \quad
\rho_j= \sqrt\frac{j+2}{2(j+1)} \,,\quad\ldots
\end{equation}
and hence: 
\begin{equation}\label{sonevals}
\rho_j \, s_1 ~\equiv~ \rho_j \, \Sfn'(\rho_j) ~=~ \frac{4}{j+2}  \,.
\end{equation}
Also note that 
\begin{equation}\label{}
\Hfn'(\rho_i)\eql\ldots\eql \Hfn^{(i)}(\rho_i)\eql 0\,,\qquad
\Hfn^{(i+1)}(\rho_i)\not=0\,.
\end{equation}
The fact that $S'(\rho_i)>0$ imposes a  strong constraint on solutions because it directly implies that if $H(\rho)\neq 1$ then at most one such zero can be present in any given solution.

It is useful to analyze the metric around the points, $\rho_j$. We define $(\tphi,\tpsi)=(\phi,\phi-3\rho_i^2 \psi)$ and the one-form $\chi= d\tpsi+A$ then find
\bea
ds_7^2 &=& \frac{1}{3(\rho_j^2 s_1)}\frac{d\rho^2}{\rho_j-\rho}  + \frac{s_1}{3}(\rho_j -\rho) \Blp d\tphi+\blp  \coeff{1}{\rho_j s_1}-1)\chi\Brp^2 \non \\
&&+ \rho_j^2 ds_{B_4}^2 +\frac{1}{6}\Blp 1 + (\rho_j-\rho )(h_1-\coeff{2}{s_1 \rho_j^2})\Brp \chi^2 + \cO((\rho_j-\rho)^2)\,.
\eea
After the change of co-ordinates $\rho_j-\rho=\coeff{1}{4}x^2$ the $(x,\tphi)$ part of the metric is
\be
ds_{(x,\tphi)}^2\sim dx^2 + \frac{\rho_j^2 s_1^2}{4}x^2 d\tphi^2  ~=~ dx^2 +x^2 \Big( \frac{2}{(j+2)} d\tphi\Big)^2 \,,
\ee
where we have used (\ref{sonevals}).  Regularity at $x=0$  means that  $ \frac{2}{(j+2)} \tphi$ must have period $2 \pi$ and since $\tphi = \phi$, we have 
\be
\label{phiperiod1}
 \phi ~\equiv~  \phi+ (j+2)\,\pi \,.
\ee
From \eq{killsp} we see that, for the supersymmetry to be well-defined, $\phi$ can be given a periodicity of 
\be
 \phi ~\equiv~  \phi+ n\,\pi  \,,\ \ \ \ n\in \ZZ\,,
\ee
and we also note that the flux defined by  (\ref{fxanzuv}) and (\ref{SEstr}) is independent of $\phi$.     Therefore the identification  (\ref{phiperiod1}) is compatible with our background and its supersymmetry.

\subsection{Expanding $\rho=0$}

Unless $U(\rho)\ra 0$ at some leftmost endpoint of $I$, the interval $I$ must terminate at $\rho=0$ because the $\CC \PP^2$ pinches off at this point.  Analytic solutions at $\rho=0$ fall into two distinct families. The first family depends on two parameters $(s,h)$ 
\begin{equation}\label{SHfam1}
\begin{split}
\Sfn(\rho) & \eql s\,\rho^4-{1\over 3}hs^3\,\rho^{10}+{1\over 6}h^2s^5\rho^{16}+{5\over 54}h^3s^7\,\rho^{22}-{1\over 70}(1-h^2)s^7\rho^{24}+\ldots \,,\\[6 pt]
\Hfn(\rho) & \eql h+{1\over 7}(1-h^2)s^4\,\rho^{14}-{2\over 15}h(1-h^2)s^6\,\rho^{20}-{1\over 11}(1-h^2)s^6\,\rho^{22}+\ldots\,.
\end{split}
\end{equation}
The second family has just one parameter $(s)$ and a ``linear behaviour'' at $\rho\rightarrow 0$,
\begin{equation}\label{SHfam2}
\begin{split}
\Sfn(\rho) & \eql  s\rho+\Big({9s^3\over 16}-{s^7\over 4}\big)\,\rho^3 +
\Big({1359s^5\over 2560}-{43 s^9\over 64}+{31 s^{13}\over 160}\Big)\,\rho^5+\ldots\,,\\[6 pt]
\Hfn(\rho) & \eql {3\over 2s^2}-\Big({9\over 4}-s^4\Big)\,\rho^2-\big({9s^2\over 32}-{5s^6\over 4}+{s^{10}\over 2}\Big)\,\rho^4+\ldots \,.
\end{split}
\end{equation}
For this family, the metric around $\rho=0$ is
\begin{equation}
ds_7^2=d\rho^2 + \rho^2 \Blp  ds_{B_2}^2 + \blp d\psi -\frac{s^2}{3} d\phi +A_2 \brp^2\Brp  + \blp \frac{1}{4 s^2} +\frac{1}{72}(12 s^4 - 45)\rho^2 \brp d\phi^2 + \cO(\rho^4)\,,
\end{equation}
so we see that for regularity the range of $\psi$ is given by
\begin{equation}
0\leq \psi \leq 2\pi \,.
\end{equation}
This is consistent with the flux \eq{AAnsatz} and the supersymmetry \eq{killsp}.

From (\ref{solXs}) we see that for this second, ``linear'' class, $U(\rho)$ goes to a constant as $\rho \to 0$  while the asymptotics (\ref{SHfam1}) implies that $U(\rho)$ diverges as $\rho \to 0$.  Thus regular solutions must have the linear asymptotics of (\ref{SHfam2}).

\subsection{Solutions with $U(\rho_0)=0$}

\begin{figure}[bth!]
\begin{center}
\includegraphics[width=3 in]{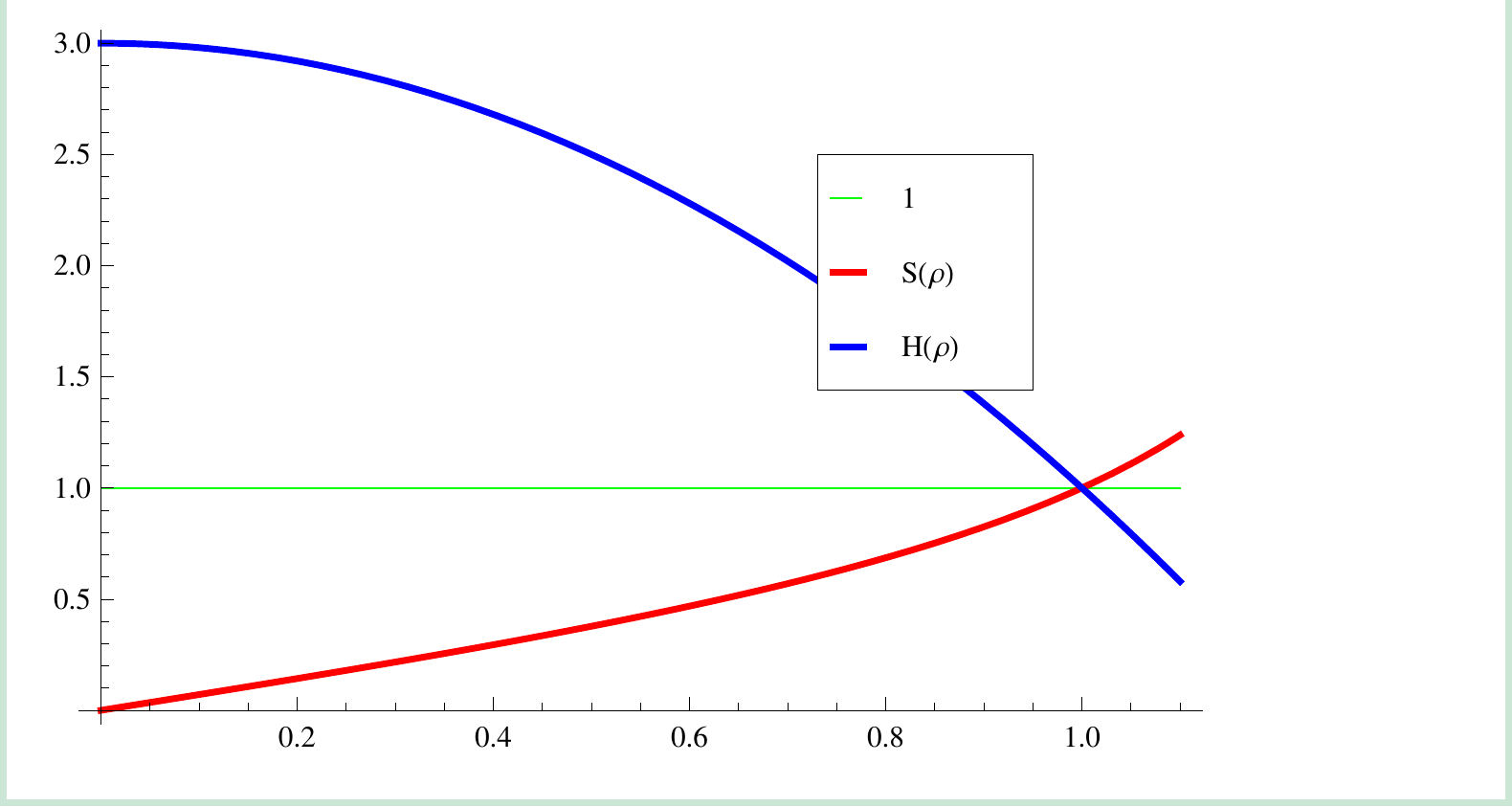}\qquad
\includegraphics[width=3 in]{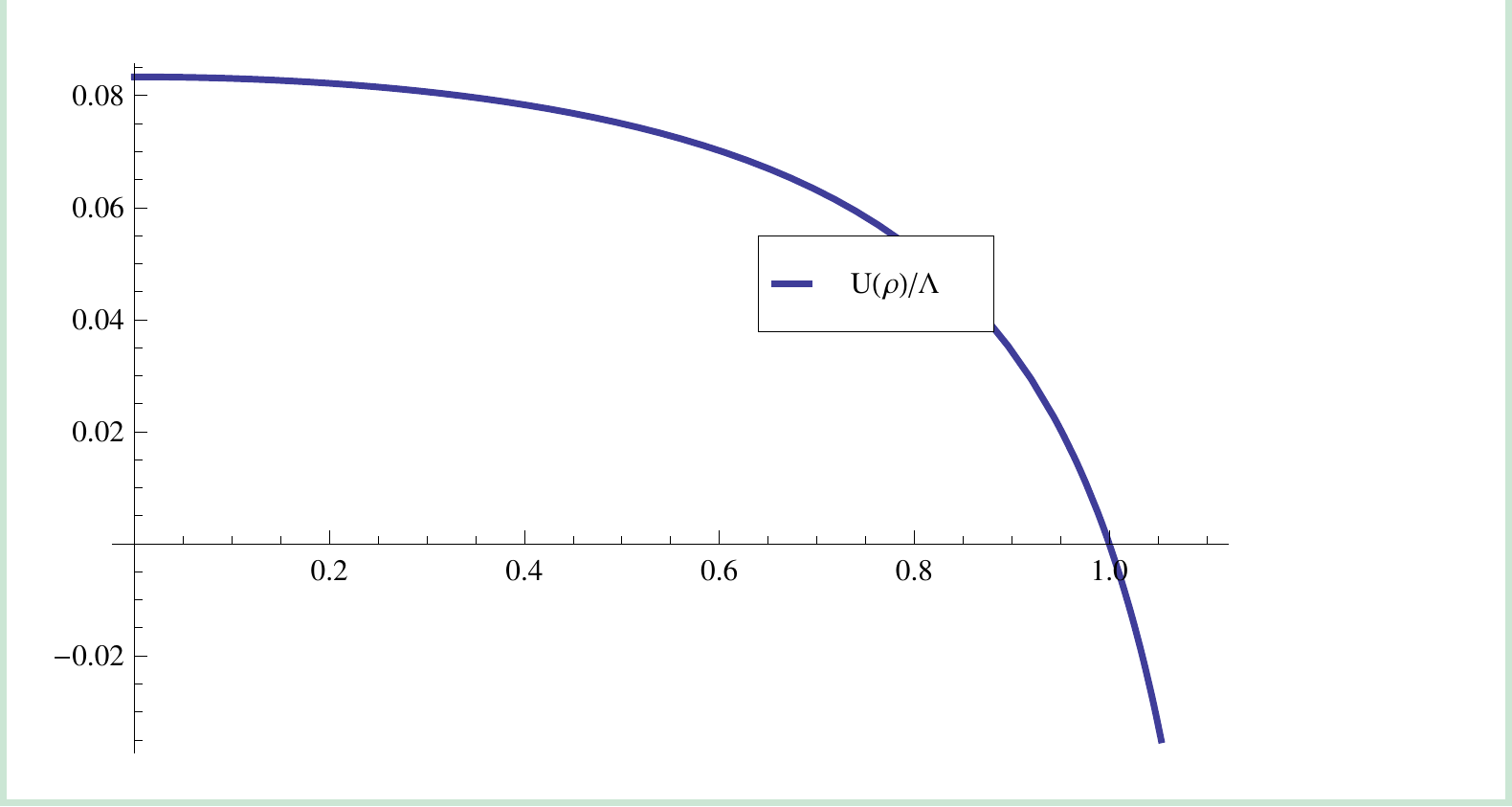}
\end{center}
\caption{\label{cpwfig}
 The CPW solution.}
\end{figure}

The CPW-solution \eq{CPWsol} is plotted in Fig~\ref{cpwfig}. Expanding the rightmost endpoint of $I$ it has $U(\rho_0)=0$ with $h_1=-4$. Around the leftmost endpoint it belongs to the second family and has $U(0)=1$ and  $s=\frac{1}{\sqrt{2}}$.

\begin{figure}[bth!]
\begin{center}
\includegraphics[width=3 in]{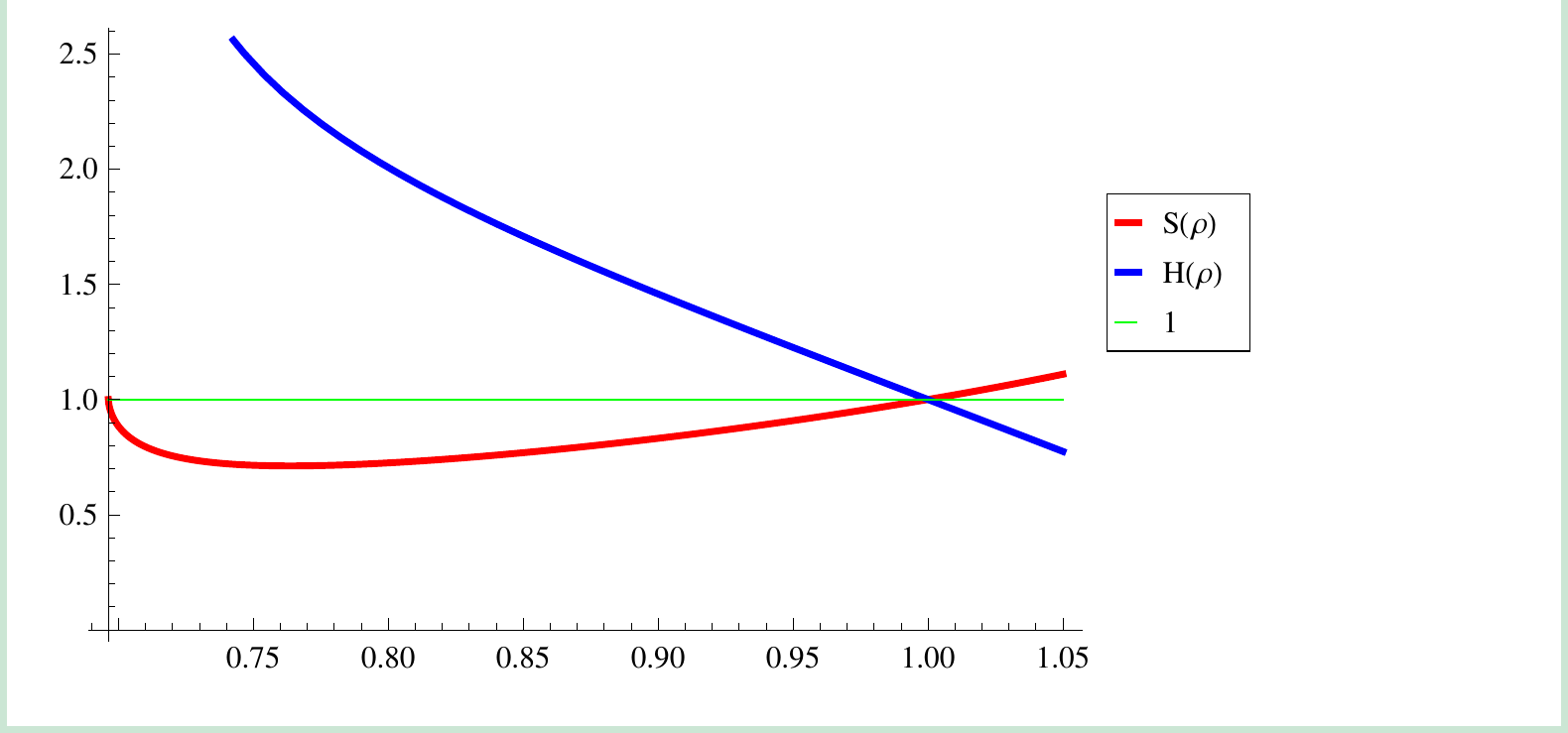}\qquad
\includegraphics[width=3 in]{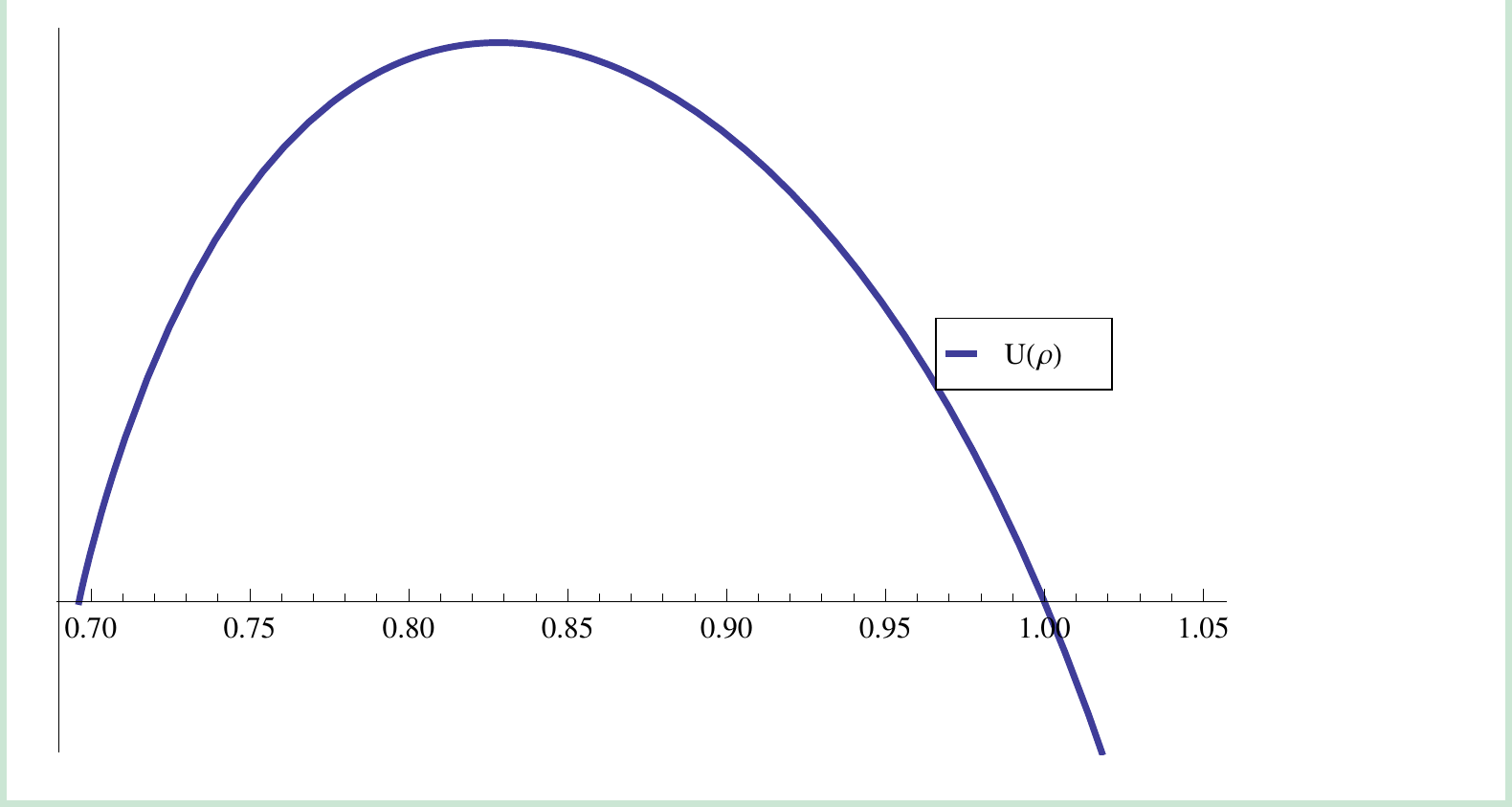}
\end{center}
\caption{\label{rho0m45}
 Solutions for $\rho_0=1$ and $h_1=-4.5$. The left zero is around $\rho=0.696287$.}
\end{figure}

We have analysed the solution space with $U(\rho_0)=0$ by varying $h_1$.  For $h_1<-4$, the solutions are similar to the one in Fig.~\ref{rho0m45} in particular they have a diverging warp factor $H(\rho)$.

\begin{figure}[bth!]
\begin{center}
\includegraphics[width=3 in]{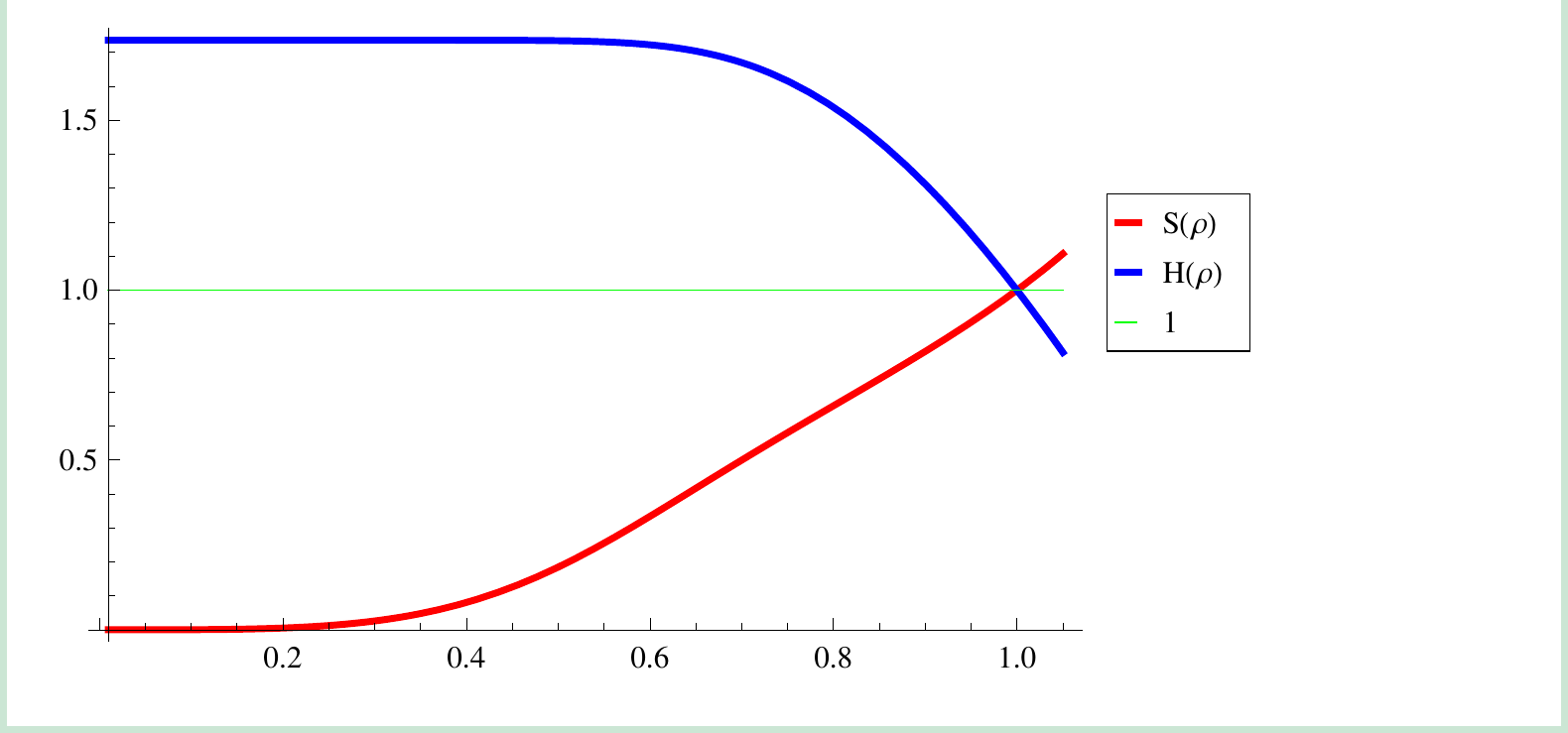}\qquad
\includegraphics[width=3 in]{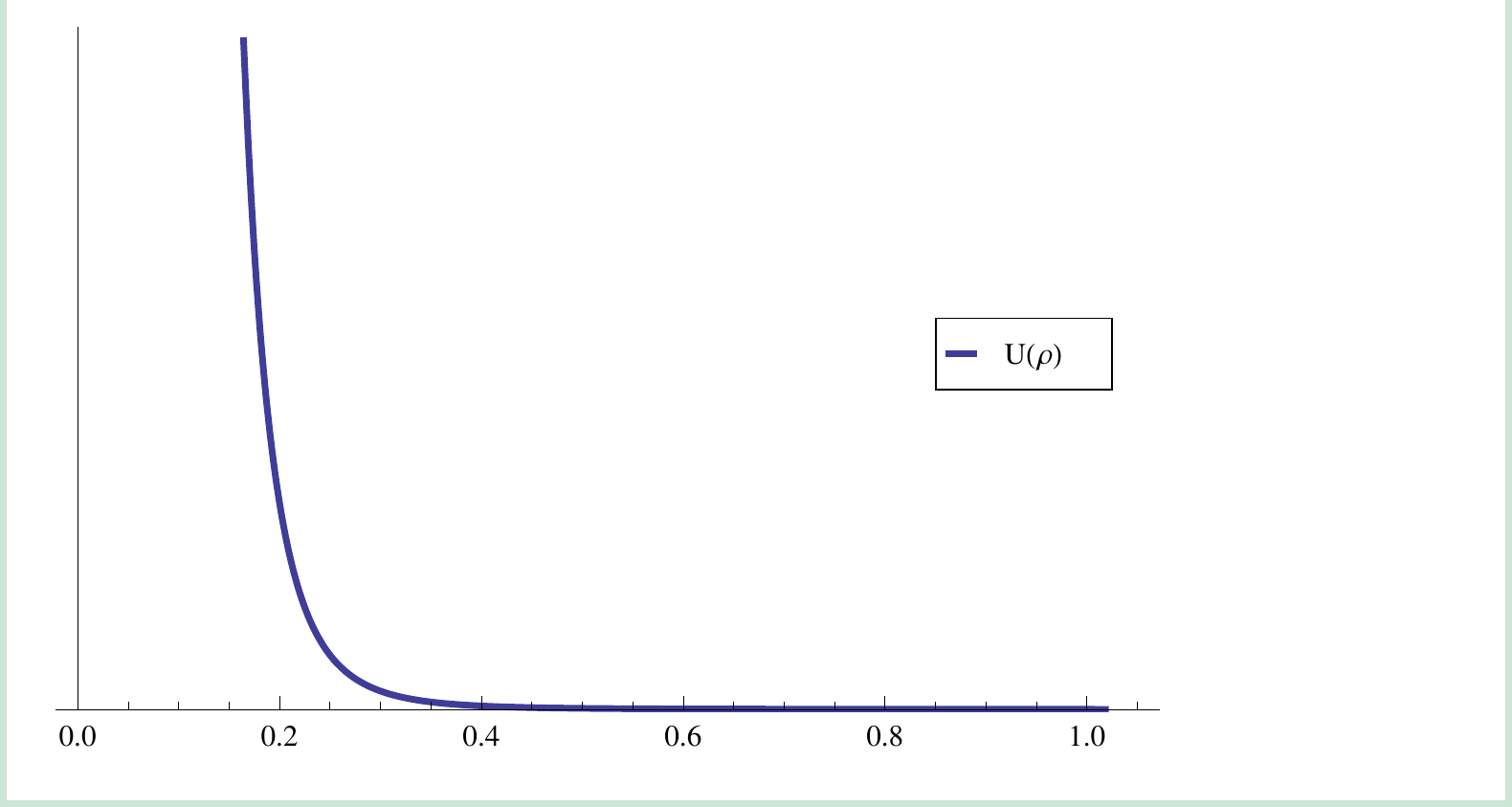}
\end{center}
\caption{\label{rho0m35}
 Solutions for $\rho_0=1$ and $h_1=-3.5$. The singular point is at $\rho=0$.}
\end{figure}

On the other side, solutions for $h_1>-4$ look qualitatively as in  Fig.~\ref{rho0m35} and have diverging $U(\rho)$.     It is clear that the CPW solution is at the interface between the two types of solutions. In particular,  $U(\rho)$ in the CPW solution has finite value at $\rho\rightarrow 0$. Also note that in this singular solution, $\Sfn$ and $\Hfn$ have the asymptotics of the form (\ref{SHfam1}), while the regular CPW solution have asymptotics (\ref{SHfam2}).

\subsection{Solutions with $U(\rho_1)=0$}
\label{goodsol}

\begin{figure}[bth!]
\begin{center}
\includegraphics[width=3 in]{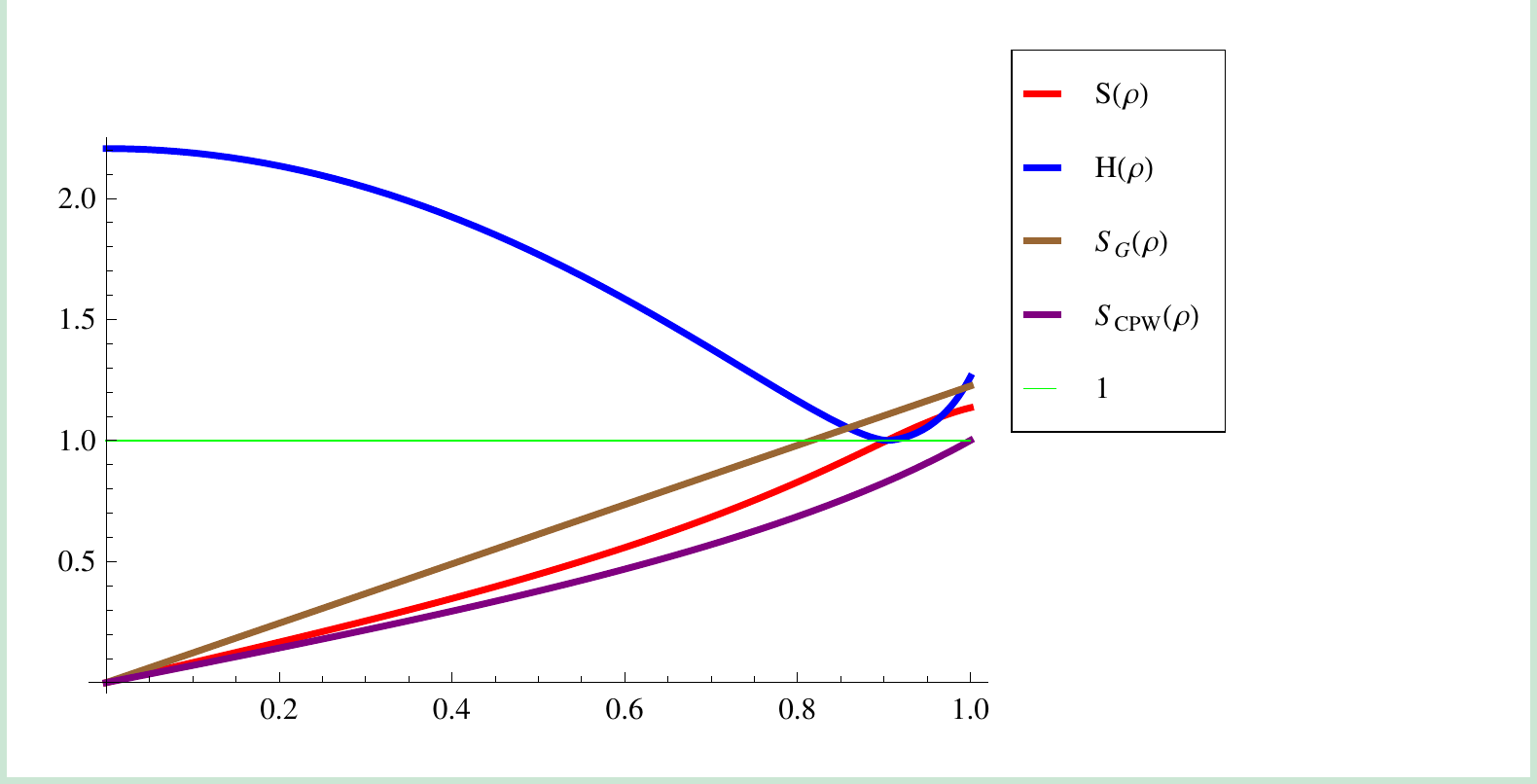}\qquad
\includegraphics[width=3 in]{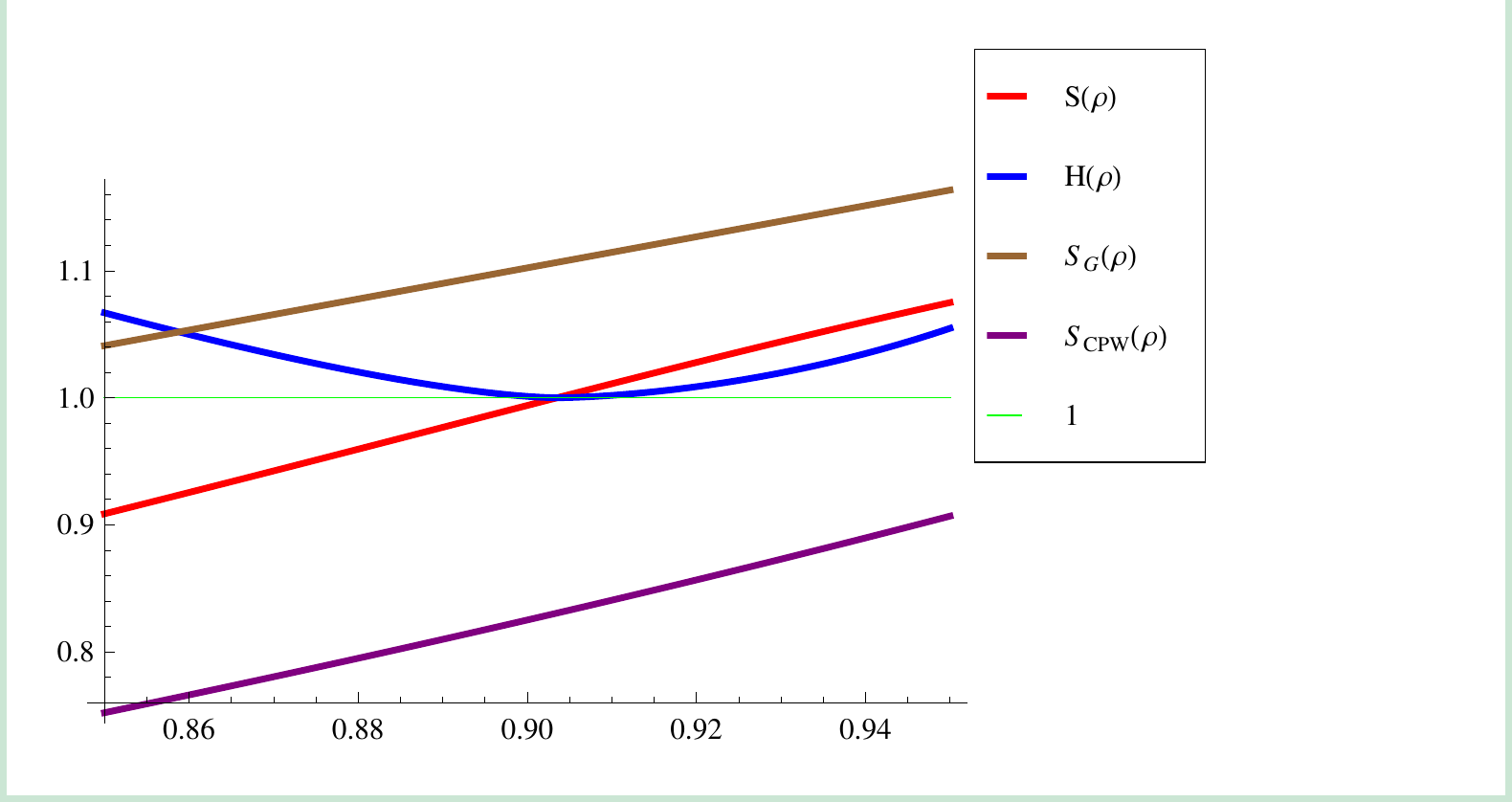}
\end{center}
\caption{\label{zeroshoot}
Shooting from $\rho=0$ with $s=0.8247$. }
\end{figure}

Here we find one new numerical solution for  $\rho \in I=[0,\rho_1]$.   This is a completely new solution and is a central result of this paper.
We obtain the solution ``shooting'' from $\rho =0$ with initial velocities determined from (\ref{SHfam2}) by a choice of the parameter $s$.  Empirically, we find that the numerical solutions seems to be singular unless $s$ is restricted to  $1/\sqrt{2}<s<\sqrt{3/2}$ and within this range we  find solutions like the ones in Fig.~\ref{zeroshoot}.  We know that the target of this shooting algorithm must be $\Hfn(\rho_1) =\Sfn(\rho_1)   =1$ at   $\rho_1 = \frac{\sqrt{3}}{2}$ and this requires delicate adjustment of the initial data.  We were not able to resolve precisely the value of $s$ that results in exactly this target data but it is evident from Fig.~\ref{zeroshoot} that $s \approx 0.8247$.

This solution has the topology of $S^7$ and is a natural generalization of the CPW solution. It would interesting to demonstrate that this solution is the IR limit of a holographic RG flow.

\subsection{Solutions with $U(\rho_2)=0$}
We also find a solution space at $\rho_2=\sqrt{2/3}$ ($h_1=h_2=0$), the solutions are parametrized by $h_3=h$. The lowest terms in the series expansion are:
\begin{equation}\label{}
\begin{split}
s_1 & \eql \sqrt{3\over 2}\,,\quad s_2\eql s_3\eql 0\,,\quad s_4\eql -{3\over 4}\sqrt{{3\over 2}}\,h\,,\quad s_5\eql -{9\over 40}\,h\,,\quad s_6\eql -{21\over 16}\sqrt{3\over 2}\,h\,,\\[6 pt]
h_1 & \eql h_2\eql 0\,,\quad h_3\eql h\,,\quad h_4\eql {3\over 2}\sqrt{3\over 2}\,h\,,\quad h_5\eql {9\over 8}\,h\,,\quad h_6\eql {1\over 32}(3\sqrt 6\,h+40\,h^2)\,. 
\end{split}
\end{equation}
One solution in this class is the analytic solution \eqref{GSeqs} for $c=0$:
 \begin{equation}\label{s7sol}
\Sfn(\rho)\eql \sqrt{3\over 2}\,\rho\,,\qquad \Hfn(\rho)\eql 1\,.
\end{equation}
which is plotted in Figs.~\ref{gntS} and \ref{gntU}.    This also belongs to the second family around $\rho=0$ with $s=\sqrt{3/2}$ while around the rightmost endpoint it has $h_k=0$. This solution is, of course, nothing other than the round $S^7$.

\begin{figure}[bth!]
\begin{center}
\includegraphics[width=3 in]{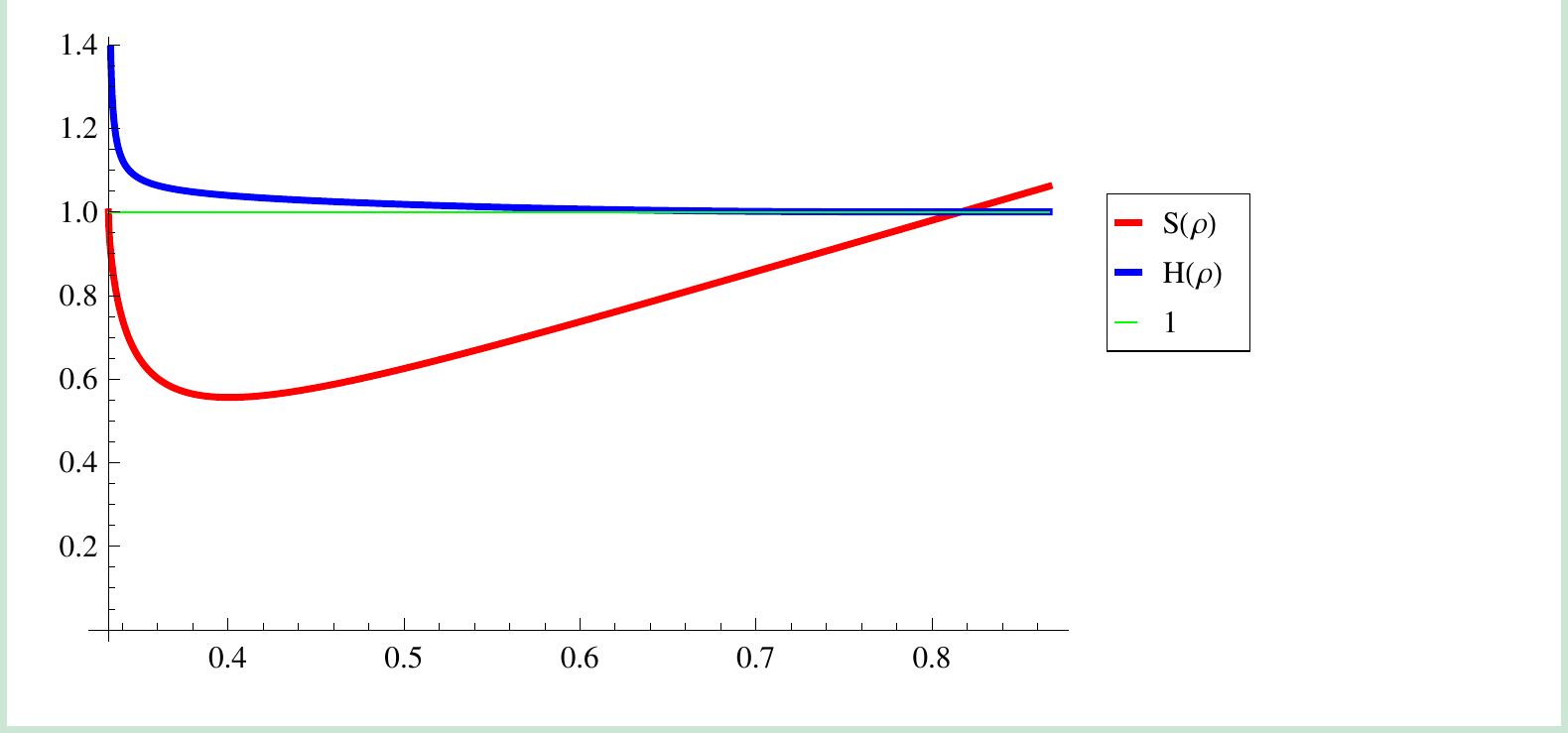}\qquad
\includegraphics[width=3 in]{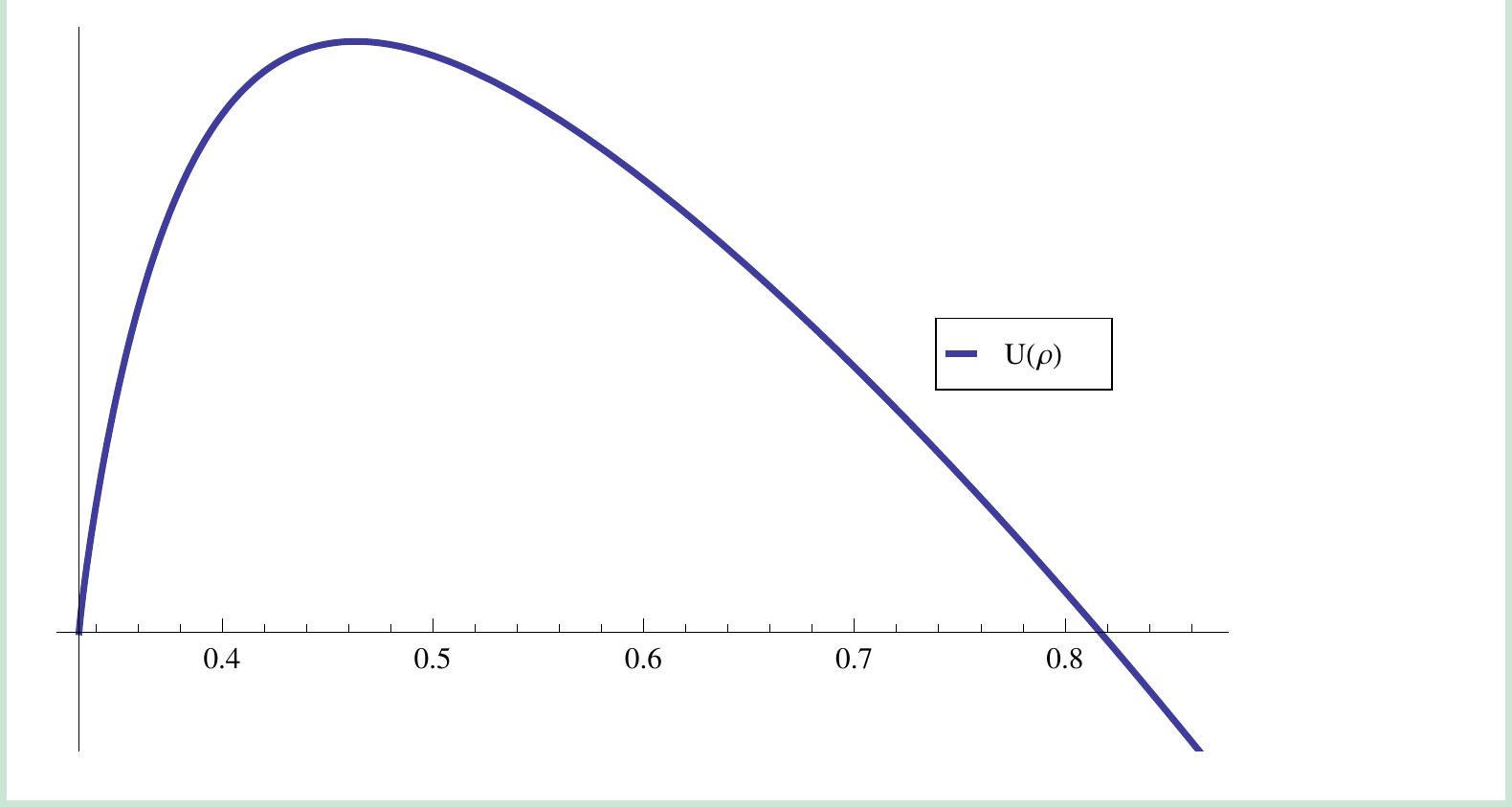}
\end{center}
\caption{\label{rho2m1}
Solutions for $\rho_2=\sqrt{2/3}$ and $h=-1$. The singular point is at $\rho=0.332142$.}
\end{figure}

\begin{figure}[bth!]
\begin{center}
\includegraphics[width=3 in]{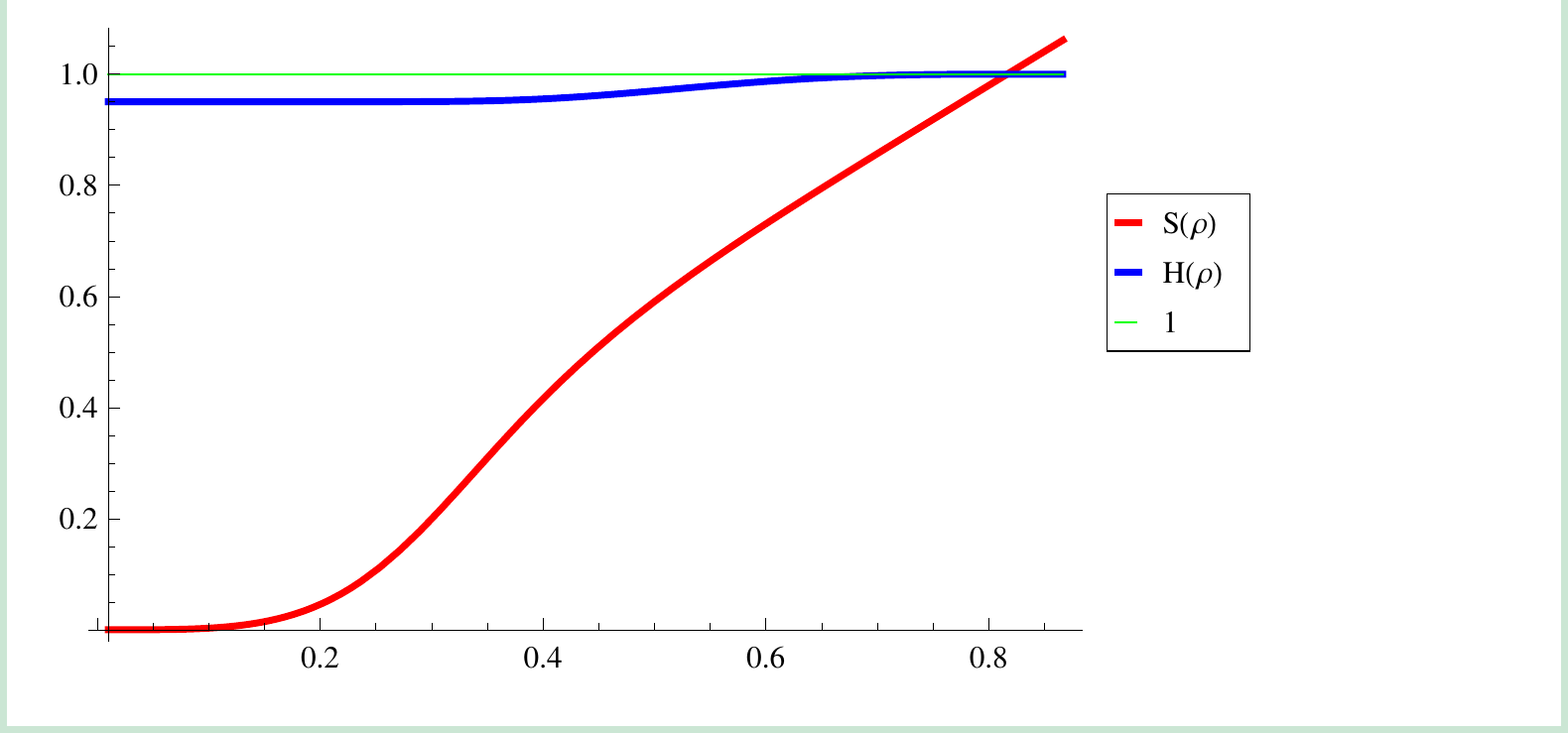}\qquad
\includegraphics[width=3 in]{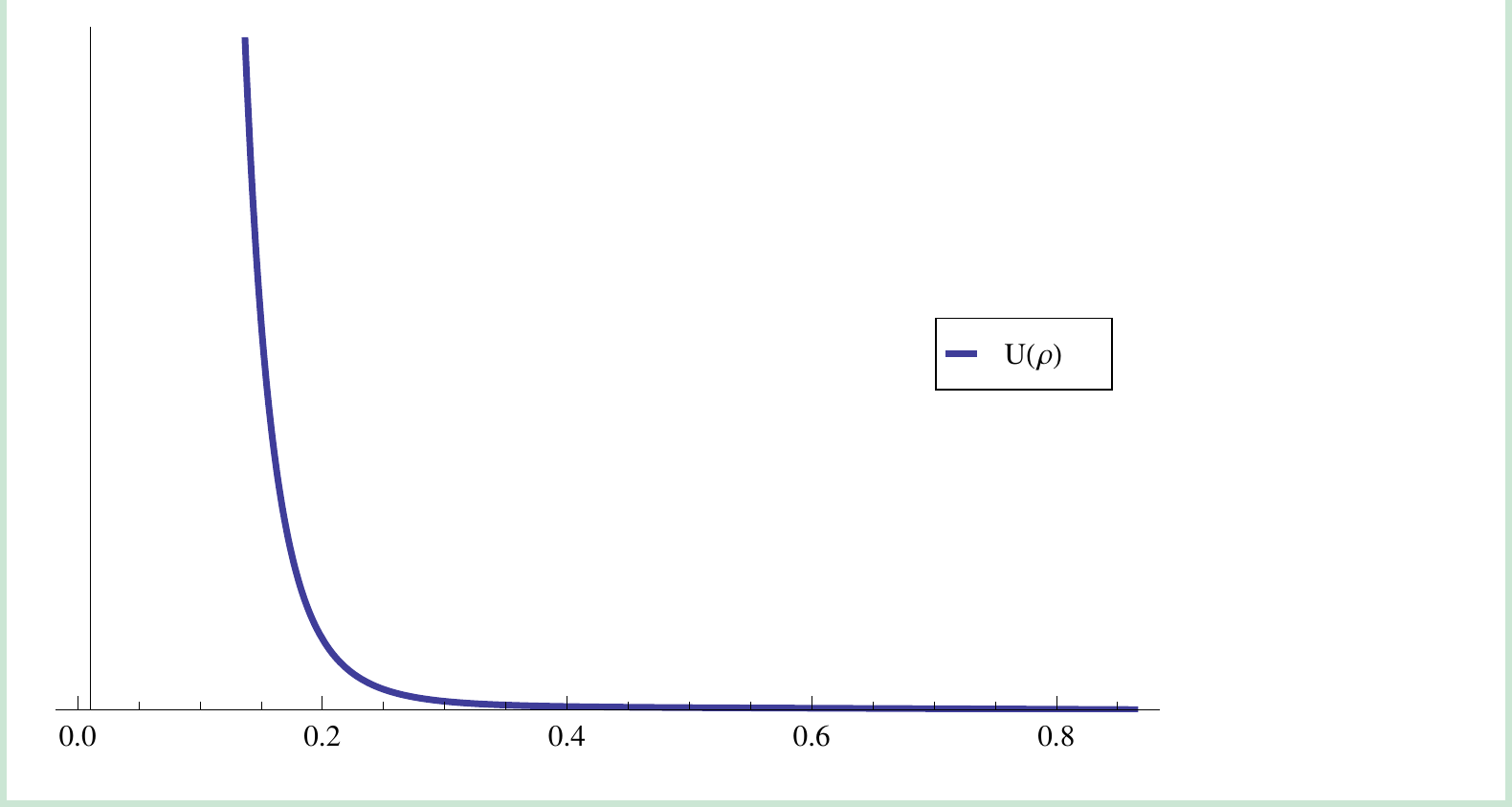}
\end{center}
\caption{\label{rho2p2}
Solutions for $\rho_2=\sqrt{2/3}$ and $h=2$. The singular point is at $\rho=0$.}
\end{figure}

We have scanned the solution space in this class and have found evidence that once more we have two generic types of singular solutions with the analytic solution lying in between. Representatives of these two classes are plotted in Fig.~\ref{rho2m1} and \ref{rho2p2} 

\subsection{Solutions with $U(\rho_k)=0,\ k\geq3$}

\begin{figure}[bth!]
\begin{center}
\includegraphics[width=3 in]{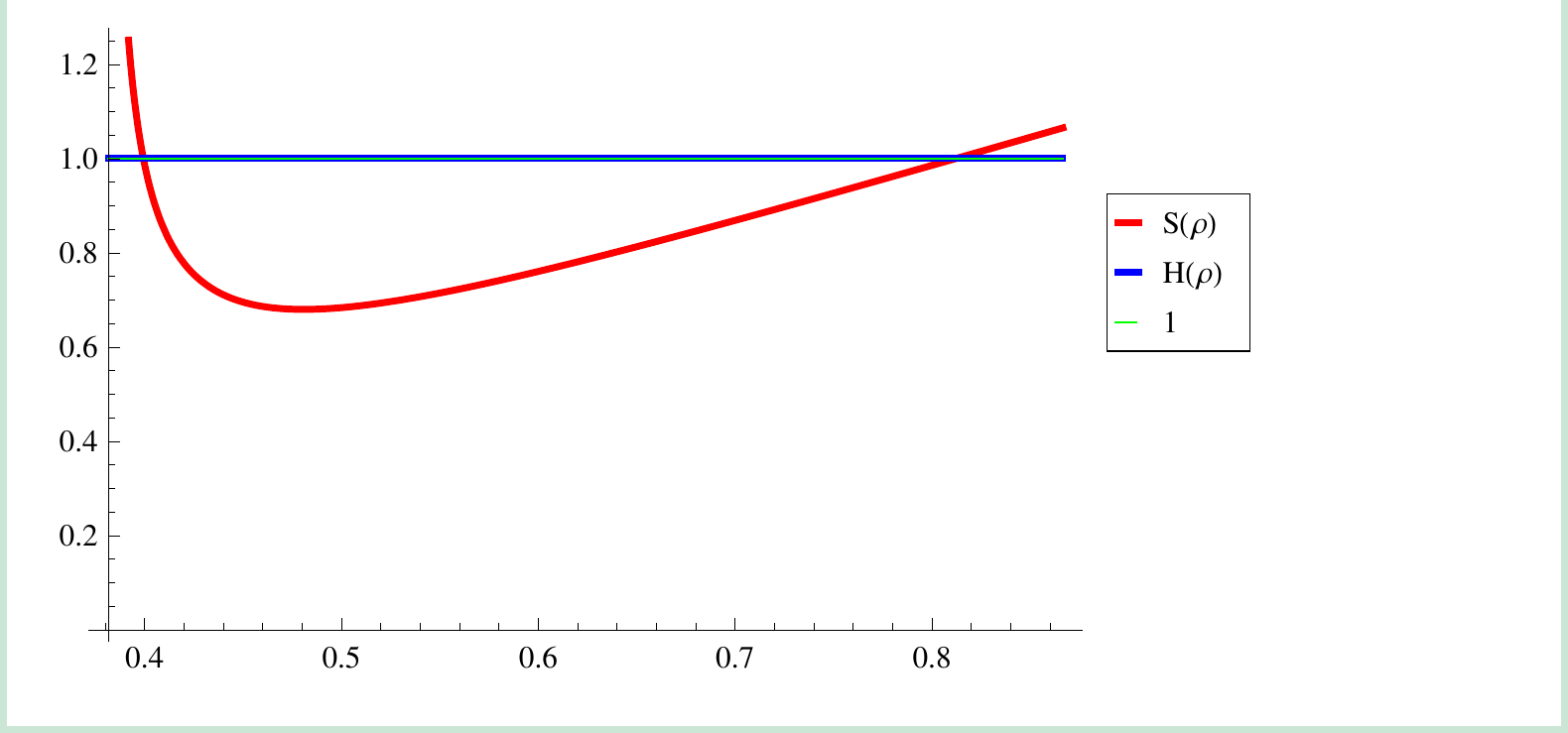}\qquad
\includegraphics[width=3 in]{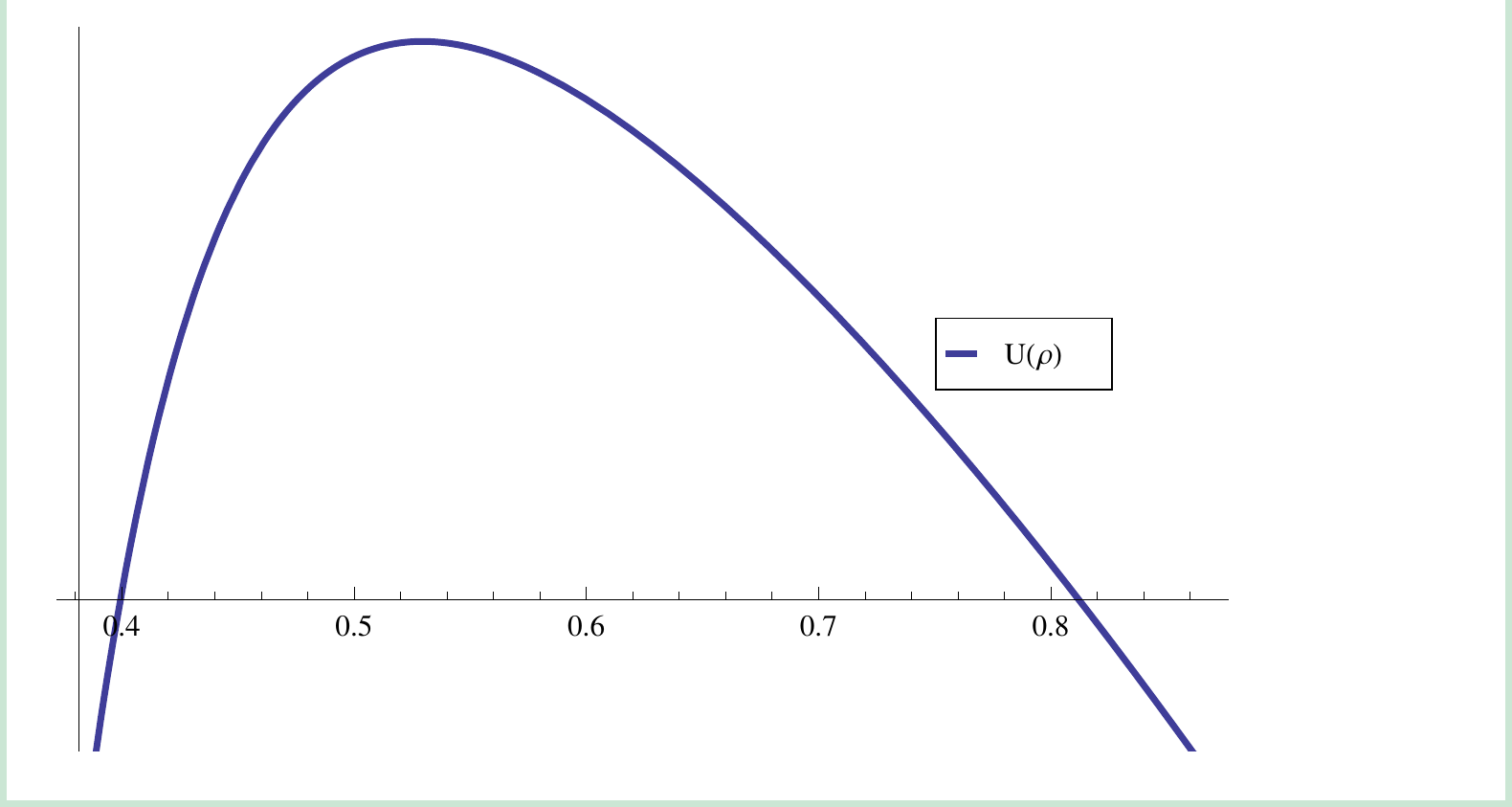}
\end{center}
\caption{\label{rho30}
Solutions for $\rho_3=\sqrt{5/2}/2$ and $h=0$. Both ends are analytic.}
\end{figure}

For $\rho_k,\ k\geq3$ solutions are parametrized by $h_{k+1}=h$, with $h_1=\ldots=h_k=0$. For each $k$, there is an analytic solution in the family \eq{GSeqs} with $h_i=0,\ \forall\, i$ (i.e. $H(\rho)=1$). These are, of course, precisely the GMSW metrics and an example is plotted in Fig. \ref{rho30} for $k=3$.

For solutions with fluxes we must have $H(\rho)\neq 1$ and the interval $I$ must continue all the way to $\rho=0$.  We have been unable to produce a numerical solution that map onto \eq{SHfam2} at $\rho=0$ but we have been able to find solutions that map onto \eq{SHfam1} however, as noted above, the latter have badly singular metrics because $U(\rho)$ diverges as $\rho=0$.
\vskip 10mm

\subsection{Summary}

After examining all the numerical solutions we find precisely one new, smooth solution:  It is the one exhibited in Section \ref{goodsol} and has non-trivial flux and has $\rho \in I=[0,\rho_1]$.  All other solutions appear to be singular.

\section{Discussion}

We have found a new $\Neql2$ supersymmetric, $AdS_4$ flux solution in M theory in which the topology of the compactification manifold is still $S^7$.  It appears as a single, isolated regular solution in a broad class that generalizes the CPW solution \cite{Corrado:2001nv}.     Given the form of our new solution and the close parallels between the CPW solution and the  $\Neql2$ supersymmetric, $AdS_5$ flux solution in IIB supergravity \cite{Pilch:2000ej, Pilch:2000fu} (the PW solution) it is natural to ask whether there is another solution to IIB supergravity that is analogous to the solution found here.   Indeed, there is  a similarity between the equations \eq{SHeqs} we have studied here and the general study of $AdS_5$ solutions of IIB supergravity in \cite{Gauntlett:2005ww}.

In fact, following \cite{Pilch:2004yg}, we have  analyzed a IIB version of the Ansatz 
\bea
ds_{10}^2&=& H^{1/2} ds_{AdS_5}^2 - L^2 H^{-1/2}  ds_{5}^2\,, \\
 ds_{5}^2&=& X_2^2 \, d\rho^2+\rho^2 ds_{\CC\PP^1}^2+\left(X_4\, \eta_{\CC\PP^1}+X_5 \, d\phi\right)^2+X_6^2\, d\phi^2\, ,
\eea
with a constant dilaton, a ``holomorphic'' $G^{(3)}$ flux as in \cite{Pilch:2004yg}, and
 \begin{equation}\label{}
F^{(5)} \eql dC^{(4)}+*dC^{(4)}\,,\qquad C^{(4)}\eql {1\over 4}H^4\cos\beta\,dx^0\wedge dx^1\wedge dx^2\wedge dx^3\,,
\end{equation}
where $\beta$ is the dielectric polarization angle.
The resulting BPS equations are\footnote{We have checked that \eq{SHIIBeqs} agrees with eqs (5.5) of \cite{Gauntlett:2005ww} after appropriate redefinitions.
}
\begin{equation}\begin{split} \label{SHIIBeqs}
{d{\Sfn}\over d\rho} & \eql (n+2)\, {\Sfn\over \rho}-{(n+2)}{\lambda\over\Lambda} \,\Hfn\,{\Sfn^3\over\rho^3}\,,\\
{d\Hfn\over d\rho} & \eql {(n+2)}{\lambda\over\Lambda}\,(1-\Hfn^2)\,{\Sfn^4\over \rho^3(1-\Sfn^2)}\,,
\end{split}
\end{equation}
with  $n=1$ and $\lambda/\Lambda=2/3$. The same equations with $n=2$ and $\lambda/\Lambda=1/2$ are the ones studied in this paper \eq{SHeqs}.
The metric functions are given by
\begin{equation}
X_2    \eql -{\sqrt \Hfn}\,{ \Sfn\over\rho\,\sqrt{1-\Sfn^2}}\,, \quad 
X_4   \eql {1\over\sqrt \Hfn}\,{\rho^2\over \Sfn}\,, \quad
X_5    \eql  {2\over 3}\,{1\over\sqrt{\Hfn}}\,\Sfn\,,\quad
X_6    \eql {2 \over 3}\,\sqrt \Hfn\,\sqrt{1-\Sfn^2}\,.
\end{equation}
Interestingly the zeros of $g_{\rho\rho}^{-1}$ are still given by \eq{diszer}
\be
\rho_j = \sqrt\frac{j+2}{2j+2}\,.
\ee
The round $S^5$ solution is
\be
S(\rho)=\rho\,,\ \ \ H(\rho)=1\,,
\ee
and solution of \cite{Pilch:2000ej} corresponds to 
\be
S(\rho)={\rho\over\sqrt{2-\rho^2/2}}\,,\ \ \ H(\rho)=2-{3\over 4}\rho^2\,.
\ee

One interesting conclusion of our study is that the solution of section \ref{goodsol} has no analogue in IIB. For the PW solution and the round $S^5$ solution in IIB, the interval $I$ ends on the right at $\rho_0$ and $\rho_1$ respectively, there is no intermediate point whereas in the M-theory solutions of this paper, the CPW point and the round point end at $\rho_0$ and $\rho_2$, leaving a new point $\rho_1$ corresponding the new solution presented here.

\vskip 15mm
\noindent {\bf Acknowledgements} NH would like to thank Michela Petrini for discussions. The work of KP and NPW was supported in part by DOE grant DE-FG03-84ER-40168.  NPW is grateful to the IPhT, CEA-Saclay for hospitality while some of this work was done.

\providecommand{\href}[2]{#2}\begingroup\raggedright\endgroup

\end{document}